\documentclass[iop,apj,twocolumn,single]{emulateapj}
\usepackage{enumitem,graphicx,apjfonts,multirow,amsmath,bm,xcolor}

\shorttitle{The Magnetic Furnace}
\shortauthors{Augustson et al.}

\newcommand{\Msun}[1]{#1\,\mathrm{M} \,\!\scriptscriptstyle \sun\!}
\newcommand{\Lsun}[1]{#1\,\mathrm{L} \,\!\scriptscriptstyle \sun\!}
\newcommand{\Rsun}[1]{#1\,\mathrm{R} \,\!\scriptscriptstyle \sun\!}
\newcommand{\Osun}[1]{#1\,\Omega \,\!\scriptscriptstyle \sun \!}

\newcommand{\mpers}[1]{#1\, \mathrm{m \,s^{-1}}}
\newcommand{\kmpers}[1]{#1\, \mathrm{km \,s^{-1}}}

\newcommand{\dgr}[1]{#1^{\circ}}
\newcommand{\orho}{\overline{\rho}}
\newcommand{\oent}{\overline{S}}
\newcommand{\oT}{\overline{T}}
\newcommand{\opre}{\overline{P}}
\newcommand{\ddr}[1]{\frac{\partial #1}{\partial r}}
\newcommand{\ddtime}[1]{\frac{\partial #1}{\partial t}}

\newcommand{\sddp}[1]{\frac{\partial #1}{\partial \phi}}

\newcommand{\rht}{\hat{\mathbf{r}}}

\newcommand{\vv}{\mathrm{\mathbf{v}}}
\newcommand{\vB}{\mathbf{B}}
\newcommand{\vJ}{\mathbf{J}}
\newcommand{\Bcr}{B_r}
\newcommand{\Bct}{B_{\theta}}
\newcommand{\Bcp}{B_{\phi}}

\newcommand{\TME}{\mathrm{T_{ME}}}
\newcommand{\PME}{\mathrm{P_{ME}}}
\newcommand{\FME}{\mathrm{F_{ME}}}

\newcommand{\TMI}{\mathrm{T_{MI}}}
\newcommand{\PMI}{\mathrm{P_{MI}}}
\newcommand{\FMB}{\mathrm{F_{MB}}}
\newcommand{\FMV}{\mathrm{F_{MV}}}

\newcommand{\TFI}{\mathrm{T_{FI}}}
\newcommand{\PFI}{\mathrm{P_{FI}}}
\newcommand{\FFI}{\mathrm{F_{FI}}}

\newcommand{\TRD}{\mathrm{T_{RD}}}
\newcommand{\PRD}{\mathrm{P_{RD}}}
\newcommand{\FRD}{\mathrm{F_{RD}}}

\newcommand{\MTF}{\langle \Bcp\rangle}
\newcommand{\MPF}{\langle \mathbf{B}_P \rangle}

\newcommand{\vcr}{\mathrm{v}_r}
\newcommand{\vct}{\mathrm{v}_{\theta}}
\newcommand{\vcp}{\mathrm{v}_{\phi}}

\newcommand{\avg}[1]{\langle #1 \rangle}
\newcommand{\bigavg}[1]{\Big\langle #1 \Big\rangle}

\newcommand{\bomega}{\boldsymbol{\Omega}}
\newcommand{\tht}{\hat{\boldsymbol{\theta}}}
\newcommand{\pht}{\hat{\boldsymbol{\phi}}}
\newcommand{\grad}{\boldsymbol{\nabla}}
\newcommand{\dvg}{\boldsymbol{\nabla}\boldsymbol{\cdot}}
\newcommand{\curl}{\boldsymbol{\nabla}\boldsymbol{\times}}
\newcommand{\cross}{\boldsymbol{\times}}
\newcommand{\cnabla}{\boldsymbol{\cdot}\boldsymbol{\nabla}}

\begin{document}

\title{The Magnetic Furnace: Intense Core Dynamos in B stars}
\author{Kyle\ C. Augustson$^{1,2,3}$, Allan\ Sacha Brun$^3$, \& Juri\ Toomre$^{1}$ }
\affil{$^{1}$JILA and Dept. of Astrophysical \& Planetary Sciences, University of Colorado, Boulder, CO 80309, USA}
\affil{$^{2}$ High Altitude Observatory, NCAR, Boulder, CO 80307-3000, USA}
\affil{$^{3}$DSM/IRFU/SAp, CEA-Saclay and UMR AIM, CEA-CNRS-Universit\'{e} Paris 7, F-91191 Gif-sur-Yvette, France}
\email{kyle.augustson@cea.fr}

\begin{abstract}
  The dynamo action achieved in the convective cores of main-sequence massive stars is explored here through 3D global
  simulations of convective core dynamos operating within a young $\Msun{10}$ B-type star, using the anelastic spherical
  harmonic (ASH) code. These simulations capture the inner 65\% of this star by radius, encompassing the convective
  nuclear-burning core (about 23\% by radius) and a portion of the overlying radiative envelope. Eight rotation rates
  are considered, ranging from 0.05\% to 16\% of the surface breakup velocity, thereby capturing both convection that barely
  senses the effects of rotation and other situations in which the Coriolis forces are prominent. The vigorous dynamo action
  realized within all of these turbulent convective cores builds magnetic fields with peak strengths exceeding a
  megagauss, with the overall magnetic energy (ME) in the faster rotators reaching super-equipartition levels compared
  to the convective kinetic energy (KE). The core convection typically involves turbulent columnar velocity structures
  roughly aligned with the rotation axis, with magnetic fields threading through these rolls and possessing complex
  linkages throughout the core.  The very strong fields are able to coexist with the flows without quenching them
  through Lorentz forces. The velocity and magnetic fields achieve such a state by being nearly co-aligned, and with
  peak magnetic islands being somewhat displaced from the fastest flows as the intricate evolution proceeds.  As the
  rotation rate is increased, the primary force balance shifts from nonlinear advection balancing Lorentz forces to a
  magnetostrophic balance between Coriolis and Lorentz forces.
\end{abstract}


\section{Introduction}

Massive O and B stars are the great recyclers in galaxies, with their short lives, strong winds, and possible
supernova endings dominating the dispersal of the products of nuclear burning.  The ultimate fate of such stars depends
upon how much mass is lost through radiatively driven winds in the course of their evolution, and this is likely
influenced by both rotation and magnetism.  If their initial masses exceed about $\Msun{8}$, they have the potential to
end their lives as core-collapse supernovae, leaving behind a neutron star or black hole as a central remnant, in
addition to the material that is processed and expelled \citep[e.g.,][]{woosley02, heger03, langer12}. Thus the term
``massive star'' is used to distinguish them from stars of lesser mass that likely end as white dwarfs.  Much is unknown
about the dynamics occurring within the interiors of these stars. During their main-sequence lifetime, there is a
convective core deep within a massive star and thin convective shells near the surface with a large radiative envelope
between them. Yet how these regions interact through instabilities, convective driving, and magnetic linkage remains
largely unclear. Aspects of this question are addressed here with global-scale 3D simulations. These simulations
examine the turbulent convection and sustained magnetic dynamo action within the core of main-sequence B-type stars of
various rotation rates. The gravity waves that the convection excites in the radiative envelope are also captured within
the same simulated domain.

As a massive B-type star, a $\Msun{10}$ zero-age main-sequence star with solar metallicity has a radius of about
$\Rsun{4}$ and a luminosity of about $\Lsun{5700}$ based on a MESA model \citep[e.g.,][]{paxton11}.  This energy output
drives fast, mass-laden winds and ultraviolet radiation from its photosphere, which strongly influences its surroundings
\citep[e.g.,][]{kudritzki00, uddoula08, owocki11, walder12}.  With a central temperature of 31.2~MK and density of 8.8
g~cm$^{-3}$ in such a star, the fusion in the core, dominated by the CNO cycle, rapidly consumes hydrogen, leading to a
main-sequence lifetime of only 21 million years. The resulting large luminosity is carried outward by a combination of
radiation and convective motions, with the vigorous convection occurring in a nearly spherical core that occupies about
half of the star by mass and 23\% by radius. The convective transport accounts for about half of the total luminosity
within the core. The convective flows are expected to be extremely turbulent and hence are inherently time-dependent and
three-dimensional. Beyond modifying the structure of the core itself, this convection also influences the dynamics of
the radiative zone above it and may impact the surface properties of the star through three primary mechanisms:
convective overshooting and excitation of gravity waves, rotational mixing processes, and magnetic linkages
\citep[e.g.,][]{claret07, brott11, maeder12}.

\subsection{Observed Rotation and Magnetism}

Observations using spectral linewidths have shown that the majority of main sequence B-type stars are fairly rapid
rotators \citep{abt02}. The average projected equatorial velocity of these stars is about $\kmpers{140}$ on the
main-sequence, but it has a large standard deviation. Indeed, observations of single stars in the Large Magellanic Cloud
and also in field stars yield a wide range of rotational velocities, from very slow rotation to about 70\% of the
breakup speed \citep[e.g.,][]{keller04, hunter08}. More recent Doppler measurements using Fourier transforms of spectral
lines have shown that for a well selected sample population there is a bimodal distribution of projected rotational
velocity \citep{dufton13}. These two populations of main-sequence B-type stars consist of a more rapidly rotating
population with an average rotational velocity of about $\kmpers{180}$ and another peaking at $\kmpers{20}$ or less.  It
has been suggested that the stars in this latter population may possess magnetic fields that have acted to spin them
down through their winds.

The rotation rates of the core and radiative envelope of some B stars, notably $\beta$-Cepheid variables, have been
estimated using asteroseismology.  While no definite trends in the relative rotation of the core and envelope have
emerged, all stars with such measurements have been slow rotators. In particular, consider \citet{aerts03}, which ruled
out the rigid rotation of a B3V star.  Moreover, it was found that the core may be rotating three times faster than the
envelope in $\nu$ Eridani \citep{pamyatnykh04}. However, \citet{briquet07} deduced that the B2 star $\theta$ Ophiuchi is
rigidly rotating.  Recent access to a larger set of gravity mode splittings using Kepler data for a slowly pulsating B8V
star suggests nearly rigid rotation of core and envelope \citep{triana15}. There is currently too little data to exclude
the possibility of radial differential rotation. Yet the very small angular velocity contrasts for at least three B-type
stars may point to an effective angular momentum coupling, possibly by gravity waves or magnetic fields during their
evolution.

Recent observational surveys by consortia such as MiMeS (Magnetism in Massive Stars) and BOB (B Fields in OB Stars) have
been directed toward measuring magnetic fields on the surfaces of massive stars, some using spectropolarimetric Zeeman
Doppler imaging techniques. They report that only about 7\% of B-type stars exhibit large-scale magnetic fields
\citep[e.g.,][]{donati09, wade11, grunhut12, morel14}. There is also indirect evidence that a larger fraction may
possess magnetic fields, but that they may be too weak or of too small a spatial scale to be detected using
spectropolarimetry.  What remains largely unknown is how these magnetic fields arise. Four complementary theories that
attempt to address the origins of these magnetic fields are: a strong core dynamo \citep{moss89, charbonneau01, brun05};
a shear-driven dynamo in the radiative zone \citep{spruit02, mullan05}; a fossil field surviving from the star's
formation stages \citep{braithwaite06}; or a bifurcation in the stability of weak and strong magnetic field
configurations in differentially rotating envelopes \citep{auriere07, lignieres14}. Some of the implications of such
differing origins for the observed surface magnetic fields are discussed in \citet{donati09} and \citet{langer12}.

\subsection{Simulating Massive Stars}

For this paper, a series of spherical 3D magnetohydrodynamic (MHD) simulations was carried out to examine the properties
of the magnetic fields that might be realized within the convective core of rotating main-sequence $\Msun{10}$ B-type
stars.  One can begin to build an intuition regarding the dynamics captured in these simulations by considering a
similar study of the convective core dynamos using the anelastic spherical harmonic (ASH) code in rotating $\Msun{2}$
A-type stars \citep{brun05}.  However, the field strengths and flow structures maintained within the core of the B stars
were uncertain given both the nonlinearity of the underlying physics and the freedom in choosing a parameterization of
the action of unresolved motions.  In particular, the convection within the cores of the more massive B stars is much
more vigorous, where the average convective velocities are ten times those seen in a typical A star
\citep{browning04}. Nevertheless, those A star simulations were able to realize nearly equipartition dynamo states, and
they too had little differential rotation that survived within the core when in the presence of magnetic fields.

Adding to the earlier work of \citet{browning04} and \citet{brun05}, a core-fossil field system was constructed by
introducing various fossil field configurations into the radiative envelope surrounding the core dynamo of an A-type
star in \citet{featherstone09}. In particular, the interaction between the preexisting core dynamo and a fossil poloidal
magnetic field allowed the dynamo to become super-equipartition, wherein the magnetic energy was roughly ten times the
kinetic energy, as might be expected from theory \citep{boyer84, sarson99}. Such behavior is in contrast to simulations
in the absence of a fossil field where these stars generated magnetic energies that were only 70\% of the convective
kinetic energy \citep{brun05}.  Peak field strengths in the super-equipartition states reach as high as 500~kG.  As will
be seen in the subsequent sections, the dynamo achieved within the more rapidly rotating B-type stars can achieve such
super-equipartition states within the core without the aid of a fossil field.

The dynamo action in the cores of massive B stars may yield magnetic fields that are strong enough to be capable of
buoyantly rising to the surface before thermalizing.  However, the buoyant structures must overcome the compositional
gradients and stiff stratification present at the core boundary \citep{macdonald04}. One mechanism that partially
circumvents this constraint, by reducing the magnetic field strength required to buoyantly rise to the surface, is
convective overshooting and penetration into the stably stratified envelope above the core.  This leads to mixing in
this region, which reduces the compositional and sub-adiabatic gradients there.  It also deposits magnetic field above
the steepest gradients, where the field may be strengthened through shearing by the differential rotation. This field
may grow to strong enough to become sufficiently buoyant to survive radiative exchange as it rises to the surface. Thus
the core dynamo might have direct bearing on the regions of strong magnetic field observed on the surfaces of some
massive stars. However, it may well be that what is detected are fossil fields from the formation stages of the massive
stars, and many favor this simpler suggestion rather than calling for contemporary dynamo action to explain the magnetic
fields detected in a small fraction of B stars.

Simulations of 3D core convection without magnetism in massive stars were first reported by \citet{kuhlen03} for a
$\Msun{15}$ main-sequence B star, and recently for a similar star by \citet{gilet13} using low Mach number modeling,
revealing the complexity of the resulting flows. Using a fully compressible code, \citet{meakin07} examined the
differences in 2D and 3D modeling of core convection in a nonrotating $\Msun{23}$ main-sequence O star, emphasizing that
3D convective flows are dominated by small plumes and eddies unlike the more laminar 2D flows. Angular momentum
transport by internal gravity waves generated at the interface between the convective core and radiative envelope in a
$\Msun{3}$ star has been investigated by \citet{rogers13} in a 2D equatorial domain for a range of rotation rates,
with the finding that the waves can be effective in such transport if strongly forced.
 
This paper has the following structure: Section \ref{sec:scaling} provides a simple scaling argument for how the core
magnetic field strength may be influenced by rotation as measured by a Rossby number. Section \ref{sec:formprob}
presents the MHD equations as solved in ASH and how the numerical experiments are formulated. Section \ref{sec:overview}
considers the general characteristics of the flows and a magnetic field structures acheived in the simulations. Section
\ref{sec:compare} contrasts the convective patterns and mean flows realized in a hydrodynamic model with those in a
companion MHD simulation. Section \ref{sec:examine} summarizes the properties of the MHD simulations for varying
rotation rates (and thus Rossby number).  Section \ref{sec:genmag} examines in detail the mechanisms involved in the
generation of magnetic field.  Section \ref{sec:feedback} analyzes the properties of the flows and magnetic field
structure that admit the remarkable existence of super-equipartition states. Section \ref{sec:conclude} reflects upon
the significance of our core dynamo solutions.

\begin{figure}[t!]
  \begin{center}
    \includegraphics[width=0.45\textwidth]{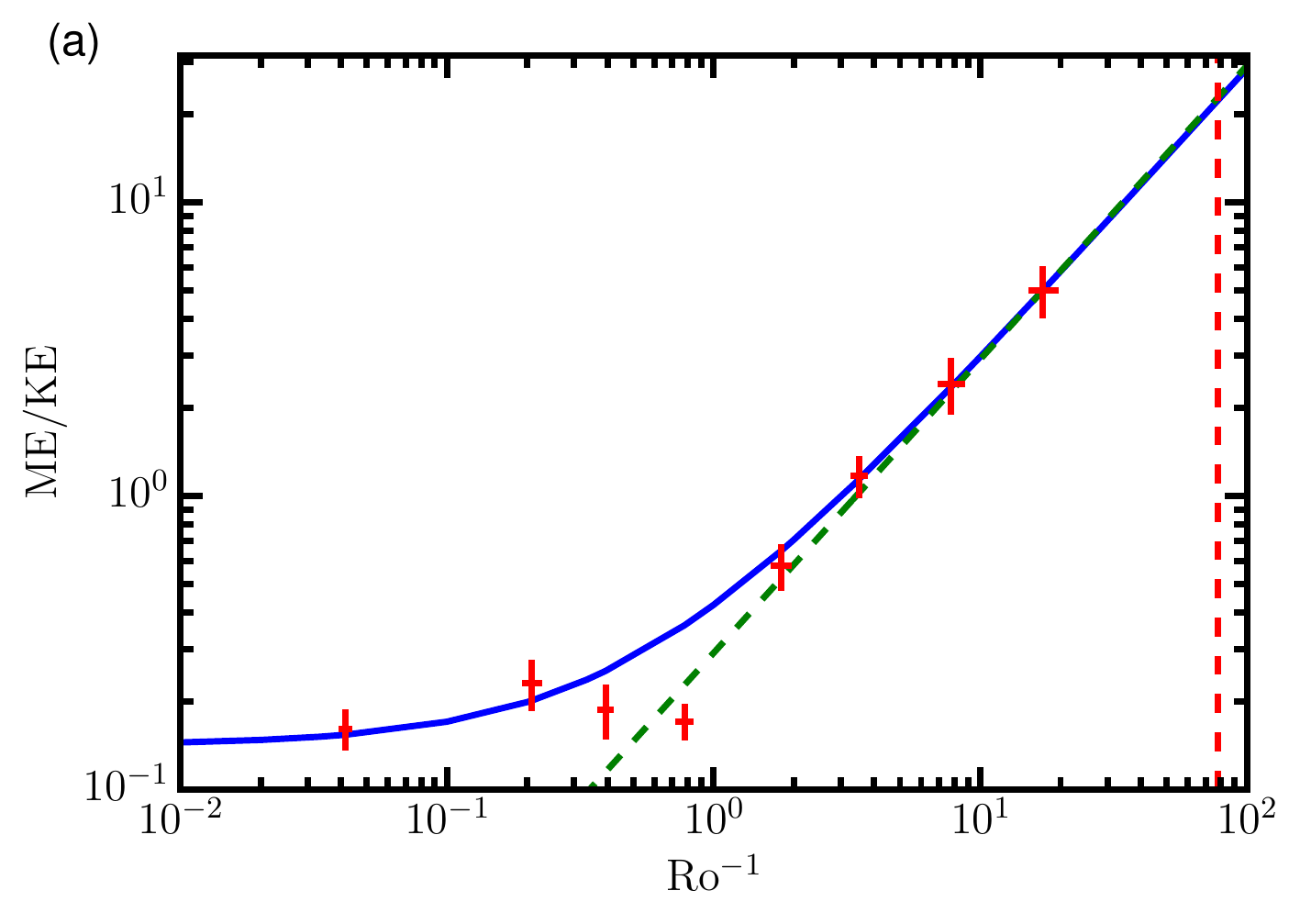}

    \figcaption{Theoretical scaling of the super-equipartition magnetic field (solid blue and dashed green lines), given
      by Equation (\ref{eqn:scaling}), with data overlain from the eight simulations examined here (red dots). These
      curves have been scaled to intersect with the last data point. The uncertainties in the data, namely its standard
      deviation in time, are indicated with red crosses. The dashed red line indicates the upper limit on the Rossby
      number that corresponds to rotating at breakup velocity. \label{fig:scaling}}
  \end{center}
\end{figure}

\section{Scaling of Magnetic and Kinetic Energies} \label{sec:scaling}

In considering the magnetic field that can be achieved in the core of massive stars, it is useful to assess the
dynamical balances that may arise in such rotating and convective systems. A relatively simple force balance involves
the Lorentz and inertial forces in a weakly rotating system that yields magnetic field strength in equipartition with
the kinetic energy contained in the convection, such as

\begin{align}
  \rho\vv\cdot\nabla\vv \approx \frac{1}{4\pi}\curl{\vB}\cross\vB , \label{eqn:forcebalance}
\end{align}

with $\vv$ the velocity, $\vB$ the magnetic field, and $\rho$ the density.  Suppose that the flow and magnetic field vary
with the length scale $L$.  Therefore, the average gradients yield an equipartition magnetic field of order

\begin{align}
  B^2_{\mathrm{eq}} \approx 4\pi\rho \mathrm{v}^2.
\end{align}

Since most of the B-type stars are rotating fairly rapidly, the core dynamo may reach a magnetostrophic state where the
Coriolis force play a key role in balancing the Lorentz force as discussed generally in \citet{christensen10} and
\citet{brun15}. First, consider that there are other forces at work in this system: pressure, buoyancy, and diffusive
forces, as was discussed in detail in the context of planetary dynamos \citep{davidson13}. If one assumes a simple model
wherein some fraction ($1-\beta$, for $0<\beta<1$) of the inertial force counterbalances those three forces, then one
arrives at a modified force balance

\begin{align}
  &\beta\rho\vv\cdot\nabla\vv + 2\rho\vv\cross\bomega_0 \approx \frac{1}{4\pi}\curl{\vB}\cross\vB, \\
  &\implies \frac{\beta}{L}\rho \mathrm{v}^2 + 2\rho \mathrm{v} \Omega_0 \approx \frac{B^2}{4\pi L},
\end{align}

\begin{align}
  &\implies \frac{B^2}{8\pi} \approx \frac{1}{2}\rho \mathrm{v}^2\left(\beta + 2 L \Omega_0/\mathrm{v}\right), \\
  &\implies \frac{\mathrm{ME}}{\mathrm{KE}} \approx \beta + Ro^{-1}, \label{eqn:scaling}
\end{align}

\noindent with $\bomega_0 = \Omega_0 \hat{e}_z$, the rotation rate of the reference frame, and with $Ro$ the Rossby
number.

The basic point is that the ratio of the total magnetic energy to the kinetic energy depends upon the degree of the
rotational constraint of the convection and upon the intrinsic ability of the convection to generate a sustained
dynamo. This can be characterized as above in Equation (\ref{eqn:scaling}), where that ratio depends upon the the inverse
Rossby number and the magnitude of the magnetic field established in a nonrotating convective core. This scaling with
and without the inertial term is shown in Figure \ref{fig:scaling}. In that figure, the blue curve is given by Equation
(\ref{eqn:scaling}) with $\beta=0.5$, whereas the green line corresponds to $\beta=0$ and leads to the scaling
$\mathrm{ME/KE} \approx Ro^{-1}$. This shows that the inertial term gives rise to a minimum magnetic energy state. It
further demonstrates that for large rotation rates the core dynamo is expected to reach super-equipartition states with
$\mathrm{ME/KE}>1$, with the possibility of achieving states with this ratio much greater than unity. The concurrence
between the measured values from the simulations and those from the above scaling relationship is examined in
\S\ref{sec:examine}.

\begin{figure}[t!]
  \begin{center}
    \includegraphics[width=0.45\textwidth]{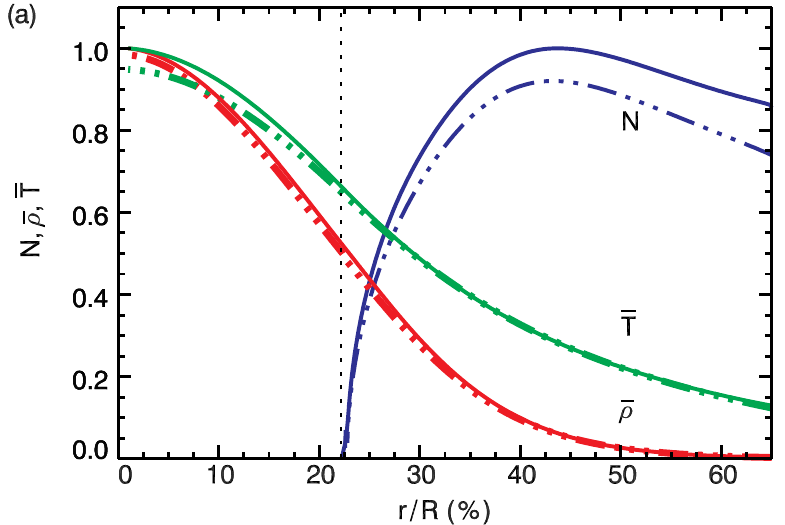} 

    \figcaption{Time-averaged mean stratification established in case {\sl M4}. Density normalized to the central value
      ($\orho$, solid red) and normalized temperature ($\oT$, solid green), with the stellar model density and
      temperature as black dashed-dotted lines. Normalized Brunt-V\"{a}is\"{a}l\"{a} frequency ($N$, solid blue), with
      stellar model values as black dashed-dotted lines. The core boundary is indicated with a dashed vertical
      line. \label{fig:backgrd}}
  \end{center}
\end{figure}

\begin{table*}[t!]
   \begin{center}
   \begin{tabular}{cccccccccccccc}
   \multicolumn{14}{c}{\bf Table 1} \\
   \multicolumn{14}{c}{Diagnostic Flow Parameters in the Four Cases} \\
   \hline
   \hline
   Case & $\mathrm{v}_{eq}/ \mathrm{v}_{breakup}$~$(\%)$ & Ra & Ta & Re & 
   Re$'$ & $\mathrm{Re_m}$ & $\mathrm{Re'_m}$ & Ro & $\Lambda_\eta$ & $\Lambda_D$  & $\kappa_{1}$ & $\tau_c$ & $\tau_{\kappa}$  \vspace{0.025truein} \\
   \hline
   {\sl M1/20} & 1/20  & $5.77\times 10^7$ & $1.66 \times 10^{5}$  &  780 & 710 & 3120 & 2840 & 24.06 & 5995 & 38.12& 8.4 &  44  (94) & 9000 \\
   {\sl M1/4}   &   1/4  & $8.17\times 10^7$ & $4.14 \times 10^{6}$  &  840 & 710 & 3360 & 2840 &  4.81 & 1936 & 11.32& 8.4 &  47 (112) & 9000 \\
   {\sl M1/2}   &   1/2  & $1.08\times 10^8$ & $1.66 \times 10^{7}$  &  920 & 770 & 3680 & 3080 &  2.54 &  923 & 4.63& 8.4 &  51 (144) & 9000 \\
   {\sl M1}       &      1  & $1.04\times 10^8$ & $6.62 \times 10^{7}$  &  860 & 830 & 3440 & 3320 &  1.29 &  389 & 2.25& 8.4 &  52 (163) & 9000 \\
   {\sl M2}       &      2  & $1.96\times 10^8$ & $2.65 \times 10^{8}$  &  600 & 570 & 2400 & 2280 &  0.56 &  263 & 2.11& 8.4 &  68 (216) & 9000 \\
   {\sl H4}       &      4  & $8.96\times 10^8$ & $1.44 \times 10^{9}$  & 1850 & 740 &  --   &   --   &  0.53 &  --  & --& 7.2 &  65  (99) & 10500 \\
   {\sl M4}       &      4  & $4.51\times 10^8$ & $1.47 \times 10^{9}$  &  580 & 560 & 2320 & 2240 &  0.28 &  209 & 1.79& 7.2 &  80 (235) & 10500 \\
   {\sl M8}       &      8  & $9.39\times 10^8$ & $8.47 \times 10^{9}$  &  530 & 520 & 2120 & 2080 &  0.13 &  132 & 1.22& 6.0 & 104 (292) & 12700 \\
   {\sl M16}     &     16 & $2.70\times 10^9$ & $5.19 \times 10^{10}$ &  520 & 510 & 2080 & 2040 &  0.06 &  102 & 0.97& 4.8 & 134 (308) & 15700 \\
   \hline
   \end{tabular}
   \end{center}
   \tablecomments{The depth of the convection zone $\mathrm{d = r_2 - r_b}$ (where $r_\mathrm{b}$ is the
     radius of the bottom of the convection zone) is the relevant length scale in the following
     parameters and is $6\times10^{10}\,\mathrm{cm}$.  
     The following diagnostic parameters are
     estimated at mid-convection zone: the Rayleigh number $\mathrm{Ra} = \Delta \oent \mathrm{g
       d^3} / \mathrm{c_P} \nu \kappa$; Taylor number $\mathrm{Ta} = 4 \Omega_{0}^{2}
     \mathrm{d}^4/\nu^2$; Reynolds number $\mathrm{Re} = \mathrm{v_{rms}} \mathrm{d} / \nu$;
     fluctuating Reynolds number $\mathrm{Re'} = \mathrm{v'_{rms}} \mathrm{d} / \nu$; magnetic
     Reynolds number $\mathrm{Re_m} = \mathrm{Pm}\mathrm{Re}$; fluctuating Reynolds number
     $\mathrm{Re'} = \mathrm{Pm}\mathrm{Re'}$; and Rossby number $\mathrm{Ro} = \overline{\omega} / 2
     \Omega_0$, where $\overline{\omega}$ is the rms vorticity.  A dissipative Elsasser number $\Lambda_{\eta} =
     \mathrm{B_{rms}^2 Pm}/\kappa_1 \rho_0 \Omega_0$ and a dynamic Elsasser number $\Lambda_D =
     \mathrm{B_{rms}^2}/(8\pi\rho_0\Omega_0\mathrm{v_{rms}}\ell_J)$ are defined, where $\ell_J$ is the typical length scale of the
     current $J$.  Thermal diffusion ($\kappa_{1}$) values are
     in units of $10^{12} \mathrm{cm^{2} s^{-1}}$. The local overturning time in the mid-convection zone,
     $\tau_c$, with the global overturning time listed parenthetically, and the thermal diffusion
     time $\tau_{\kappa}$ across the convective core are all quoted in days.}
\end{table*}

\section{Formulating the Problem} \label{sec:formprob}

The research tool employed here for modeling dynamics within stellar interiors is the 3D ASH simulation code. ASH is a
global large-eddy simulation (LES) code that solves the anelastic MHD equations of motion in a rotating spherical shell
using a pseudo-spectral method \citep{brun04}. Simulations using ASH capture the entire spherical shell geometry and
allow for global connectivity of magnetic structures and turbulent flows.  ASH employs spherical harmonic expansions
$Y^m_{\ell}(\theta,\phi)$ in the horizontal directions of the entropy, magnetic fields, pressure, and mass flux, and
fourth-order nonuniform finite differences in the radial direction to resolve the radial derivatives. The solenoidality
of the mass flux and magnetic vector fields is achieved through a streamfunction formalism. The spherical harmonic
expansion is truncated at degree $\ell_{\rm max}$, with all azimuthal orders $m$ retained in the triangular truncation,
thus yielding uniform resolution over spherical surfaces. For all the simulation here, $\ell_{\rm max} = 340$,
corresponding to 512 mesh points in the latitudinal direction ($N_{\theta}$), with the longitudinal mesh having
$N_{\phi} = 2 N_{\theta}$ or 1024 mesh points, and in radius $N_r = 500$ mesh points. The time evolution is computed
using a semi-implicit second-order Crank-Nicolson scheme for the linear terms and an explicit second-order
Adams-Bashforth scheme for the advective, Coriolis, and Lorentz terms.  The substantial computational demands for the
extended evolution required to study core dynamos are met with ASH optimized to run very efficiently on massively
parallel supercomputers.

ASH simulations cannot capture all of the very wide spectrum of scales of motion and magnetic fields expected in the
highly turbulent cores. Thus the unresolved sub-grid-scale dynamics is parameterized as effective momentum, thermal, and
magnetic diffusivities. Further, the anelastic approximation is employed to retain stratification effects and
Alfv\'{e}n waves, while filtering out sound waves, which have short periods relative to the dynamical time scales of most
interest. For B-type stars, the sound speed in the core is of the order of $\kmpers{1000}$, while the convective flows
are about $\kmpers{0.5}$. Thus the anelastic approximation is ideally suited for simulating the interior of these
stars. The Courant-Friedrichs-Lewy condition on the time step thus depends either on the local subsonic flow velocity,
the speed of Alfv\'{e}n waves or resolving the Brunt-V\"{a}is\"{a}l\"{a} frequency, rather than the local speed of
sound.

The turbulence achieved in these models is still quite removed from the intensely turbulent conditions likely present in
actual stellar convection zones. Yet with ASH, simulations of convection and dynamo action in stars of many spectral
types have made significant contact with observations and led to new ways of thinking about stellar dynamos. In
particular, ASH has been used to simulate a wide range of stars: from fully convective M dwarfs \citep{browning08}, to
Sun-like stars \citep{brown11,matt11,nelson13} and F-type stars \citep{augustson13}, and extending to the convective
core of an early A-type star \citep{brun05,featherstone09}.

\subsection{Anelastic Equations}\label{sec:asheqnsect}

The version of the ASH code utilized here implements the Lantz-Braginsky-Roberts co-density formulation
\citep{lantz92,braginsky95}, so that gravity waves are properly captured in the stably stratified, radiative envelope
\citep{brown12}. The anelastic MHD equations evolved in ASH are still fully nonlinear in the momentum and magnetic
variables \citep{brun04}. The thermodynamic variables of entropy, pressure, temperature, and density ($S$, $P$, $T$, and
$\rho$) are linearized about a spherically symmetric and hydrostatic background state as $\oent$, $\opre$, $\oT$, and
$\orho$, which are functions of the radial coordinate only.  The equation of state for this background state is a
calorically perfect gas. The equations solved in ASH retain physical units, are in spherical coordinates
$(r,\theta,\phi)$, and are evolved in time $t$. These equations are

\begin{align}
  \displaystyle \partial\vv/\partial t &= -\vv \cnabla \vv -\grad \varpi + S \mathrm{c_P^{-1}}\mathbf{g} - \Gamma\rht + 2 \vv \cross \bomega_0 \label{eqn:ashmom} \\
  \mbox{} &  + \left(4\pi\orho\right)^{-1} \left(\curl\vB\right)\cross\vB + \orho^{\,-1}\dvg{\mathcal{D}}, \nonumber \\
  \displaystyle \partial \vB/\partial t &= \curl\left[\vv\cross\vB-\eta\curl\vB\right], \label{eqn:ashind} \\
  \displaystyle \partial S/\partial t   &= -\vv \cnabla \left(\oent+S \right)
  + \left(\orho\oT\right)^{-1}\left[\nabla \cdot \mathbf{q} + \Phi + \epsilon\right], \label{eqn:asherg}
\end{align}

\noindent with $\vv=\vcr\rht+\vct\tht+\vcp\pht$, $\vB=\Bcr\rht+\Bct\tht+\Bcp\pht$, and the co-density
$\varpi = P\orho^{\,-1}$.  The solenoidality of the mass flux ($\dvg{\orho\vv}=0$) and the magnetic field
($\dvg{\vB}=0$) are maintained through the use of a streamfunction formalism. The angular velocity of the rotating frame
is $\bomega_0=\Omega_0\hat{\mathbf{z}}$.  The gravitational acceleration is $\mathbf{g} = -g(r)\rht$. In Equation
(\ref{eqn:ashmom}), $\Gamma\left(r\right)$ is the radial gradient of the mean turbulent pressure that maintains
hydrostatic equilibrium as the spherically symmetric state evolves. In the induction equation, $\eta$ is the magnetic
resistivity.  The volumetric heating arising from the combination of the CNO and PP nuclear reaction chains are
parameterized in $\epsilon$, as in \S\ref{sec:numerics}. The energy flux $\mathbf{q}$ comprises a spherically symmetric,
diffusive radiation flux and a turbulent entropy diffusion flux,

\begin{figure}[t!]
  \begin{center}
    \includegraphics[width=0.45\textwidth]{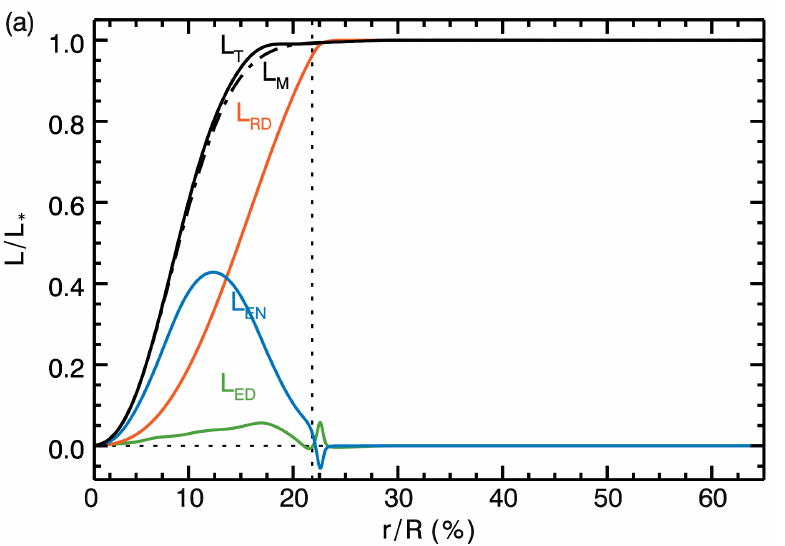} 

    \figcaption{Horizontally and time-averaged energy flux balance in case {\sl M4}. The energy fluxes are shown as
      luminosities ($L = 4\pi r^2 F$) for the total flux ($L$, black), radiative flux ($L_\mathrm{RD}$, red), enthalpy
      flux ($L_\mathrm{EN}$, blue), and the flux due to unresolved motions ($L_\mathrm{ED}$, green), as effectively
      defined in \citet{brun05}. The dashed-dotted line shows the luminosity from the input MESA model
      ($L_\mathrm{M}$). The vertical dashed line indicates the core boundary, and the horizontal dashed line shows the
      zero ordinate. \label{fig:energy}}
  \end{center}
\end{figure}

\begin{align}
  \displaystyle \mathbf{q} &= -\kappa_r \orho \mathrm{c_P} \grad \oT - \kappa \orho \oT \grad S, \label{eqn:ediff}
\end{align}

\noindent where $\kappa_r$ is the radiation diffusion coefficient and $\mathrm{c_P}$ the specific heat at constant
pressure.  The entropy diffusion flux has an effective thermal diffusivity $\kappa$ acting on the entropy fluctuations.
The viscous diffusion tensor $\mathcal{D}$ and the diffusive heating $\Phi$ are

\begin{align}
  \displaystyle \mathcal{D}_{ij} &= 2 \orho \nu \left[ e_{ij} - 1/3 \left(\dvg{\vv}\right)\delta_{ij} \right], \label{eqn:vsstress} \\
  \displaystyle \Phi &= 2\orho\nu\left[e_{ij} e_{ij} - 1/3 \left(\dvg{\vv}\right)^2\right] + \left(4\pi\right)^{-1}\eta\left(\curl\vB\right)^2, \label{eqn:heating}
\end{align}

\noindent Here $e_{ij}$ is the stress tensor and $\nu$ is the effective kinematic viscosity. The use of this set of
anelastic MHD equations requires 14 boundary conditions in order to be well posed. The following impenetrable, perfectly
conducting, torque-free, and thermal flux-transmitting boundary conditions have been adopted at the inner ($r_1$) and
outer ($r_2$) radii of the simulation:

\begin{align}
  \Bcr &= \ddr{}\left(r\Bct\right) = \ddr{}\left(r\Bcp\right) = 0, \\
  \vcr &=\ddr{}\left(\frac{\vct}{r}\right) = \ddr{}\left(\frac{\vcp}{r}\right) = 0, \\
  \ddr{S} &= 0. \label{eqn:bdrycond}
\end{align}

\subsection{Experimental Configuration}\label{sec:numerics}

These simulations model the inner 65\% by radius of a young main-sequence B-type star of $\Msun{10}$ at eight rotation
rates corresponding to 0.05\%, 0.25\%, 0.5\%, 1\%, 2\%, 4\%, 8\%, and 16\% of the surface breakup velocity, which is
$\kmpers{700}$. The name of the case corresponds to those values as in Tables 1, 2, and 3. The percentages of breakup
velocity are equivalent to 0.047, 0.235, 0.47, 0.94, 1.88, 3.75, 7.50, and $\Osun{15.0}$. The luminosity at the surface
of the star is $\Lsun{5700}$. The model of the star consists of a convective core that occupies the inner 30\% of the
computational domain by radius, and an overlying radiative exterior that extends to the outer radius of the domain. To
avoid the small time steps required when including the coordinate singularity at the origin, the innermost 1\% of the
core has been excluded.

\begin{figure}[t!]
  \begin{center}
    \includegraphics[width=0.375\textwidth]{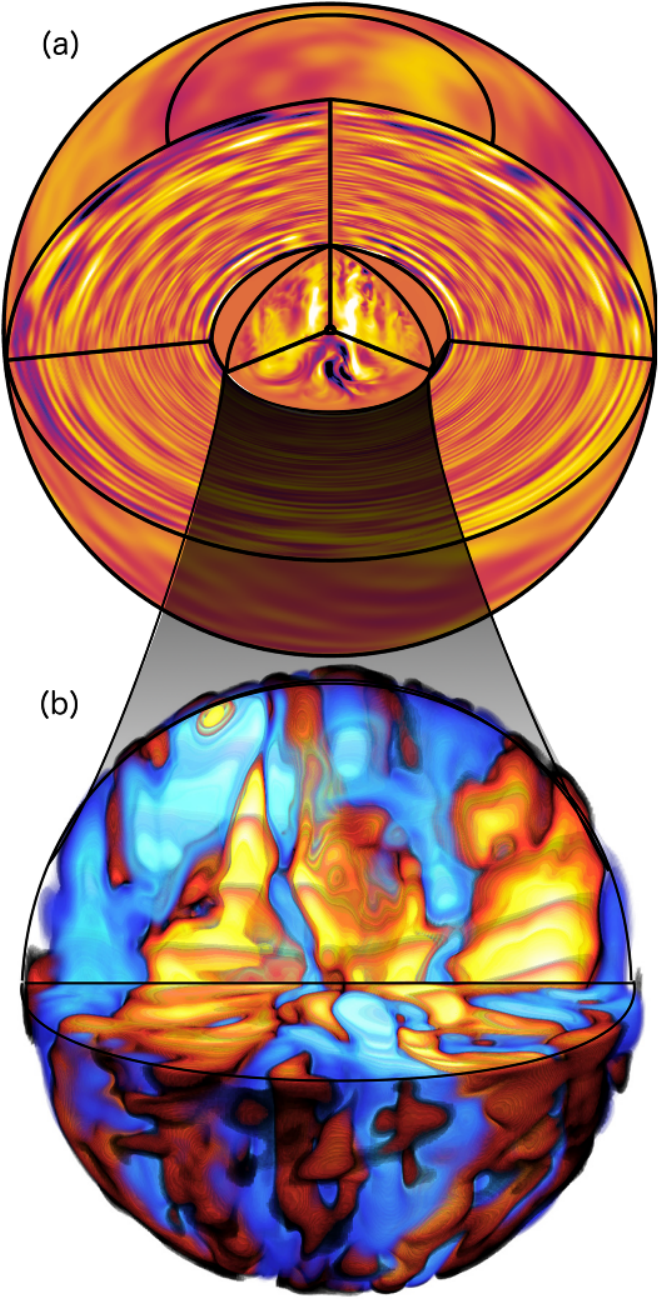} 

    \figcaption{Orthographic volume projections of the radial velocities in case {\sl M4}, with upflows as bright tones
      and downflows as dark tones. (a) The inner sphere shows the core, which has the color table scaled to
      $\mpers{100}$, and the outer sphere which captures a portion of the radiative envelope, with color table there
      scaled to $\mpers{0.1}$. (b) Radial velocities in an expanded view of only the core with one quadrant removed.
      Here the radial inflows are yellow and the outflows are blue, scaled as above. \label{fig:vrortho}}
  \end{center}
\end{figure}

\begin{figure*}[t!]
  \begin{center}
    \includegraphics[width=0.9\textwidth]{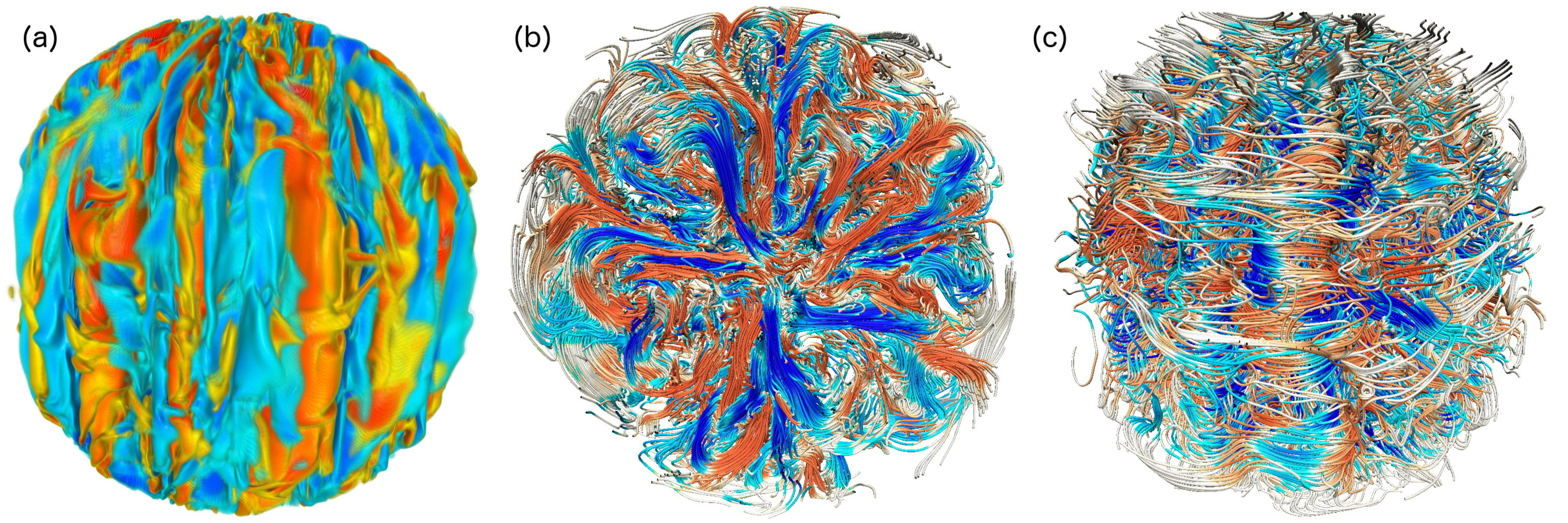}

    \figcaption{A comparison of magnetic and velocity fields in case {\sl M16}. (a) Volume rendering of the radial
      velocity in a snapshot of case {\sl M16}, showing the convective core only with the rotation axis pointing
      vertically. (b) Magnetic field line tracing, colored by the magnitude of the radial magnetic field, looking down
      the rotation axis at the equatorial cross section of the magnetic field structures. The peak field has been
      clipped to $\pm 1$~MG (red for magnetic field pointing radially outward and blue for field pointing radially
      inward) and the weakest fields about are 1~G (gray).  The strongest magnetic fields are the most interconnected,
      threading throughout the core. (c) Magnetic field line tracing seen from the same angle as in (a) and the same
      color table as in (b), showing the magnetic field of the full core. \label{fig:bsfields}}
  \end{center}
\end{figure*}

The stratification is taken from a MESA stellar evolution model that had evolved a $\Msun{10}$ star from the pre-main
sequence to about 1~Myr, or just past the beginning of its main-sequence life. The MESA code employs realistic
microphysics and employs the mixing-length approximation to the convective dynamics within the core
\citep{paxton11}. The mean values of $\partial \ln{\epsilon}/\partial\ln{T}$ and
$\partial \ln{\epsilon}/\partial\ln{\rho}$ in the core of the 1D stellar model are utilized to build a volumetric
heating in ASH that approximates a spherically symmetric nuclear burning, which gives
$\epsilon = \epsilon_0 \orho^2 \oT_{6}^{15.33}$ and
$\epsilon_0 = 2.31 \times 10^{-20} \, \mathrm{erg \, cm^{3} g^{-2} s^{-1}}$. The initial conditions for the
hydrodynamical case {\sl H4} are random entropy perturbations, and for its magnetic counterpart {\sl M4} a weak dipolar
field was introduced that was confined to the core. Most of the other magnetic cases have been obtained by either
spinning up or down the matured case {\sl M4} and continuing the simulations over at least tens of convective
overturning times to further reach a statistically stationary state.  Given the major computational resources required,
the complex evolution of these 3D solutions was monitored with care to assess the maturity of the dynamo states
achieved, and further extended as deemed appropriate. It was not feasible to extend the simulations over multiple
magnetic or thermal diffusion times across the core. However, the cases run and analyzed here appear to be a good
representation of the dynamo states that may be realized in these systems, for no statistically significant changes over
the later temporal evolution have been seen in any of the cases.

A comparison between the 1D stellar model and the background state used in our 3D simulation is shown in Figure
\ref{fig:backgrd}.  The small discrepancy in the Brunt-V\"{a}is\"{a}l\"{a} frequency $N$ displayed in Figure
\ref{fig:backgrd} arises from our use of the equation of state of a calorically perfect gas.  $\Gamma_1$ and
$\mathrm{c_P}$ are required to be constant with radius, and that the logarithmic density and temperature to be as well
fit as possible given that constraint. The radial energy flux balance is shown in Figure \ref{fig:energy}, with the
convective core and radiative envelope clearly apparent.  Deep in the core, the flows must carry much more energy given
that both the opacity is too large and the temperature gradient too small to allow radiative diffusion to carry the
local stellar luminosity. This is where convection transports much of the fusion-generated energy outward with the aid
of the radiation field. The flow deeper in the core thus arranges itself to be more well correlated with the temperature
field, which enhances the buoyancy-driving and leads to faster flows. Moving radially outward into the radiative zone,
the radiative transport of energy has risen sufficiently to carry the full luminosity of the star, which has reached its
final surface amplitude.  In these simulations, a diffusive LES energy transport mechanism, which in principle arises
from the unresolved scales of motion, carries at most 5\% of the total energy as evidenced by the solid green curve in
Figure \ref{fig:energy}. The small influence of the LES diffusion is consistent with the relatively large Rayleigh and
Reynolds numbers of these simulations, as seen in Table 1. The kinetic energy and Poynting fluxes are less than 0.1\% of
the total flux and so they are indistinguishable from zero with the linear vertical scale used in Figure
\ref{fig:energy}. The small region of overshooting just beyond the edge of the core is also of note because it serves as
a region of magnetic field storage and gravity wave excitation as will be discussed later on.

The effective diffusivities $\nu$, $\eta$, and $\kappa$ vary in radius only. The diffusion is greatly reduced in the
stable region, where motions have a small amplitude and the expected sub-grid-scale motions are subsequently smaller. In
the convective core, the Prandtl number is fixed at $\mathrm{Pr} = \nu/\kappa = 1/4$, so
$\nu=\mathrm{Pr}\,\kappa$. Similarly, the Roberts number is constant in the core with
$\mathrm{Rb} = \eta/\kappa = 1/16$, leaving $\eta = \mathrm{Rb}\,\kappa$. Outside of the core, the diffusion is tapered
to profiles that give $\mathrm{Pr}=\mathrm{Rb}=1$. The radial profile of the thermal diffusivity $\kappa$ is given by

\begin{align}
  \displaystyle \kappa\left(r\right) &= \left(\orho/\orho_{1}\right)^{-1/2}
  \left[\kappa_{1}\left(1+\exp{\left[\delta\left(r-r_{\kappa}\right)\right]}\right)^{-1} + \kappa_{2}\right], \label{eqn:kap}
\end{align}

\noindent where $\orho_{1}$ is the background density value at the lower boundary and $\kappa_{1}$ is the thermal
diffusivity at the lower boundary. The values of $\kappa_1$ are shown in Table 1 for each simulation, along with many
other diagnostic flow parameters from the middle of the convective core. The floor value of the diffusivity is
$\kappa_{2}$. The steepness of the transition between $\kappa_1$ and $\kappa_2$ is controlled by $\delta$ and occurs at
the radius $r_{\kappa}= 6.7\times10^{10}\,\mathrm{cm}$, which is just outside the core. The minimum levels of diffusion
are equal with $\kappa_{2}=\nu_{2}=\eta_{2} = 4\times10^8 \,\mathrm{cm^2 s^{-1}}$. However, the magnitude of the
molecular value of the diffusivities is still orders of magnitude less than those that can be achieved in global-scale
simulations. Nevertheless, using this set of diffusivities, the hierarchy of diffusivities is consistent with those
given by computing the molecular diffusivities following \citet{braginskii65}, which requires that
$\mathrm{Rb}<\mathrm{Pr}\ll\mathrm{Pm}$. In particular, the magnetic Prandtl number of these LESs is very close to the
molecular magnetic Prandtl number, which is roughly 10 throughout the core and drops to unity in the radiative
envelope. Thus, these stars and other stars with convective cores represent an interesting crossroads between
large-$\mathrm{Pm}$ dynamo theory and stellar astrophysics.

\begin{figure*}[t!]
  \begin{center}
    \includegraphics[width=0.9\textwidth]{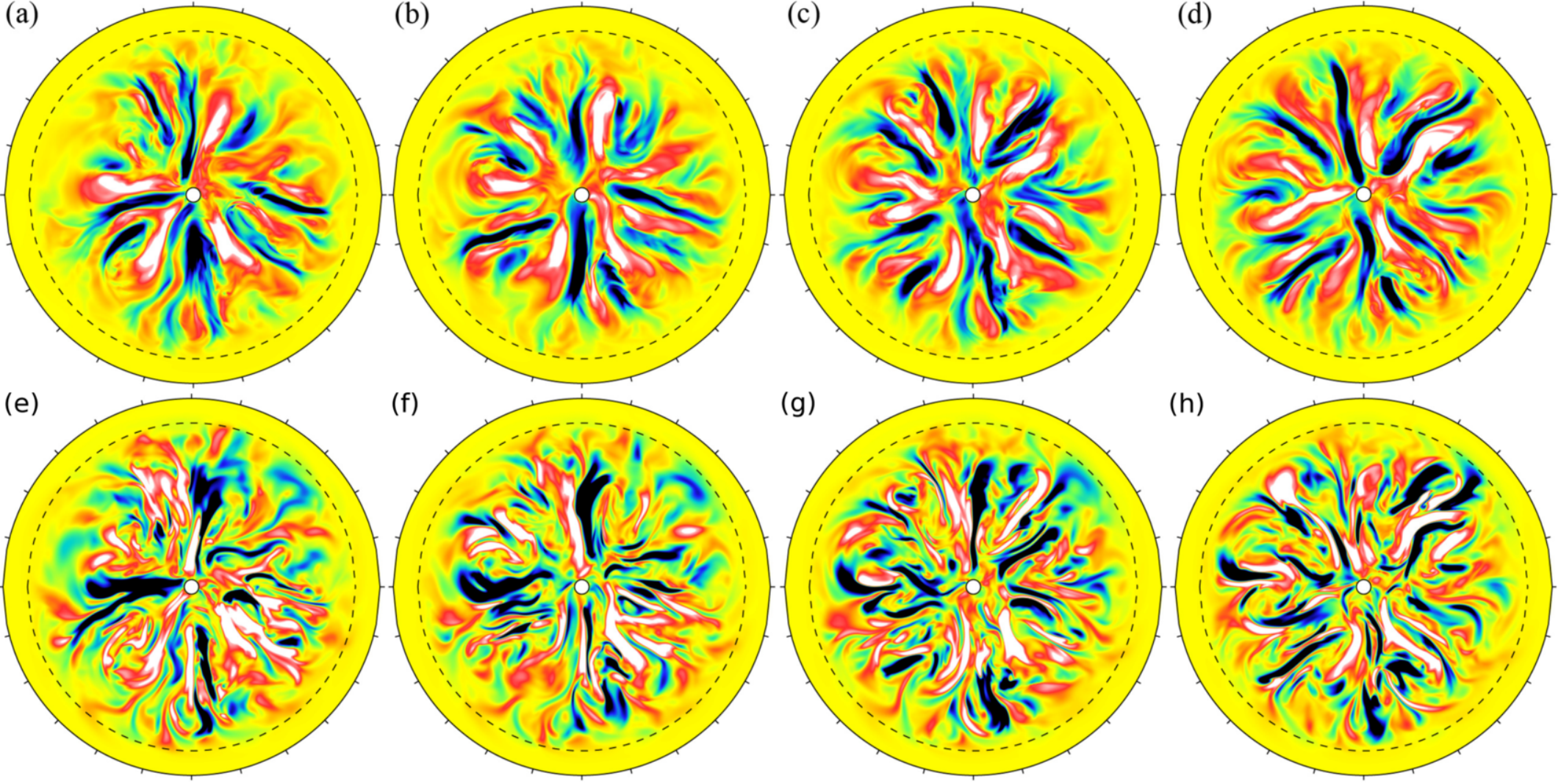} 

    \figcaption{Time evolution of radial velocity and radial magnetic field in case {\sl M16} shown in a set of
      equatorial cuts covering the convective core (delineated by the dashed circle) and a small portion of the
      radiative envelope.  The tiny central circle is the inner domain cutout in our computational domain.  (a)-(d)
      display four successive samplings of radial velocity $\vcr$ that are 20 days apart, with downflows blue and
      upflows red, and (e)-(h) show the accompanying radial magnetic fields $\Bcr$, with inwardly directed magnetic
      field in blue and outwardly pointing field in red. This interval is in the latest stage of the simulation where
      the dynamo has reached a statistically steady state.  Clipping values for radial velocity and magnetic field are
      respectively $\pm\mpers{100}$ and $\pm 10^5$~G. \label{fig:evolve}}
  \end{center}
\end{figure*}

\section{Overview of Core Dynamics} \label{sec:overview}

The large-scale convection within the cores exhibits no particular orientation at slow rotation (cases {\sl M1/2} or
slower), with the exception that the flows are radially aligned due to buoyancy forces. In contrast, faster rotation
brings with it increasingly cylindrical alignment of the convective structures with the rotation axis.  Such alignment
is consistent with the motions being more constrained by the Coriolis force, hence becoming quasi-2D in the spirit of
the Taylor-Proudman theorem \citep{pedlosky87}.  In particular, the rotationally aligned vertical structures present in
the more rapidly rotating cases are akin to the Taylor columns expected for rapidly rotating convection, in that they
are nearly invariant in the vertical direction.  These columnar structures may be quite intricate, yet have some sense
of continuity across the full core.  Their number increases with faster rotation, so that in case {\sl M16} roughly
eight to ten evolving vortical rolls are evident at any given time.  Indeed, following the work of \citet{browning04}
and \citet{miesch06}, it can be seen that the flows in the more rapidly rotating cases are quasi-geostrophic.

The flows in the core generate internal gravity waves that then propagate following complex spiral-like ray paths
through the radiative envelope above it \citep[e.g.,][]{zahn97,rogers13,alvan15}.  Such convective columns formed in the
core and the gravity waves that they drive are apparent in Figure \ref{fig:energy}(a), where an orthographic projection
of the radial velocity in the domain is shown for case {\sl M4}. Since most of the gravity waves excited here have small
radial wavelengths, they damp fairly quickly due to a relatively large thermal diffusion \citep{zahn97}.  A
self-consistent simulation of such internal gravity waves that may be pertinent to observations requires the capture of
at least the entire resonant cavity encompassing the full radiative envelope, which is not the case here. A detailed
analysis of the gravity wave spectrum generated in this set of simulations will not be presented here. The amplitude of
the flows has two scales in Figure \ref{fig:energy}(a), in the core the color table is clipped to $\mpers{\pm 100}$ in
the core and to $\mpers{\pm 0.1}$ in the radiative zone to emphasize the differing characteristics of the flow in the
convectively unstable and stable regions. Such columnar flow features are further emphasized in Figure
\ref{fig:energy}(b), which shows a volume rendering with a cutout of the radial velocity just in the core.

Turning now to the most rapidly rotating case {\sl M16}, the structure of the radial velocity field is shown in Figure
\ref{fig:bsfields}(a), and its time evolution in Figures \ref{fig:evolve}(a)-(d).  The columnar flow structures involve
fast flows interspersed with slower portions, thereby leading to a rich assembly of evolving folded sheets and tubules
with substructures.  This is evidence of some aspect of turbulent flow, yet they possess large-scale ordering that
imprints into the induced magnetic fields. Figures \ref{fig:bsfields}(b) and (c) show the magnetic field lines that
accompany this instant in time, presenting both an equatorial cut through the full core and a perspective view from just
outside the core boundary. The equatorial rendering reveals that the strong magnetic fields extend across the core and
show wrapping at larger radii consistent with the swirling sense of columnar flow cells viewed along their main
axis. The choice here to color-code the field lines by the polarity of their radial magnetic field component leads to a
change in color from blue to red or vice versa as any given field line crosses the center of the core.  The magnetic
field amplitude reaches up to 2 {\rm MG} as reported in Table 2.  The helicoidal wrapping of the magnetic field lines
close to the core-envelope boundary is evident in panel (c) for which the field has been rendered at the same instant
and in the same orientation as in panel (a).  The distinctive presence of the underlying columnar convection can be seen
in these magnetic field line structures.  However, this picture is somewhat confused by a number of magnetic field lines
connecting across many cells.  As the core edge is approached, the magnetic field topology tends to become more toroidal
in character, with the field lines wrapping around the core in the longitudinal direction. Some of these field line
characteristics are sampled in panel (c).

\begin{figure*}[t!]
  \begin{center}
    \includegraphics[width=\textwidth]{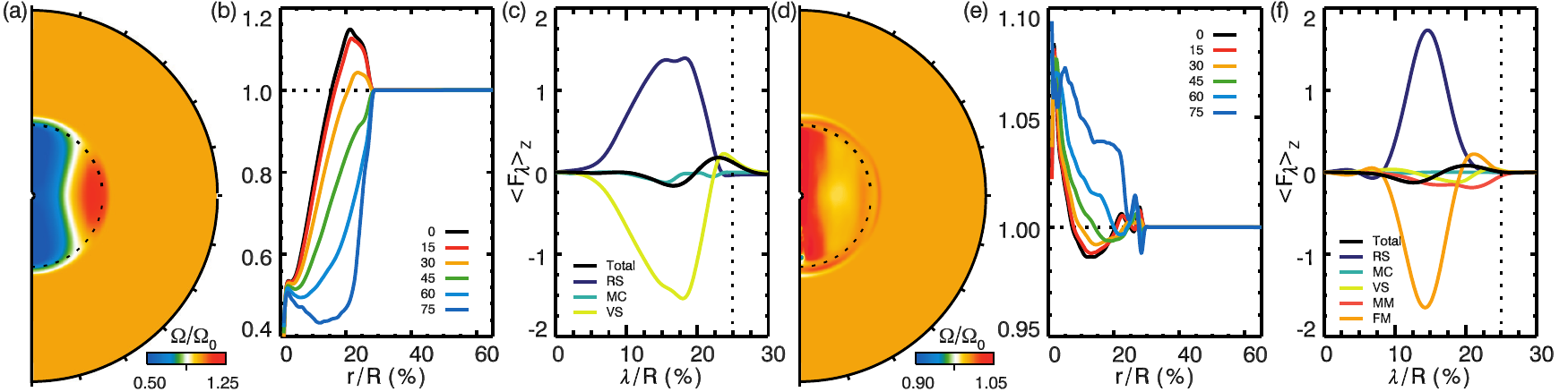}

    \figcaption{Comparison of the hydrodynamic case {\sl H4} in panels (a)-(c) and MHD case {\sl M4} in (d)-(f). (a),
      (d) Time and azimuthally averaged angular velocity $\Omega$ shown with radius and latitude. (b), (e) Radial
      dependence of $\Omega$ at specific latitudes. The vertically averaged angular momentum fluxes
      $\langle F_{\lambda}\rangle_z$ are shown with proportional cylindrical radius $\lambda/R$ for (c) case {\sl H4}
      and (f) case {\sl M4}, in units of $10^{40}\, \mathrm{g\, s^{-2}}$. The flux components are the total flux
      (black), the Reynolds stress (RS; blue), the advection by the meridional circulation (MC; teal), viscous stresses
      (VS; yellow), mean Maxwell stresses (MM; red), and fluctuating Maxwell stresses (FM; orange). The location of the
      edge of the core is indicated with dashed lines. \label{fig:hdvmhd}}
  \end{center}
\end{figure*}

The time evolution of the flow structures associated with the strongest upflows and downflows in the core occurs
approximately over a local convective overturning time, which is about 134~days for case {\sl M16}. This can be
appreciated in Figures \ref{fig:evolve}(a)-(d), where four temporally sequential snapshots of the radial velocity are
shown in the equatorial plane. These snapshots are 20~days apart, showing the coherence of radial flow structures over
roughly a convective overturning time. Since the core is rotating fairly rapidly, these flows have a distinct vortical
structure that is largely aligned with the rotation axis. Those columnar flows primarily exist in the outer region of
the core, with increasingly radially symmetric structures forming deeper in the core. The columnar flows in the outer
convection zone evolve over the shorter time scale of about 70~days. The more easily identifiable columnar structures
evolve nearly in place as the mean flows are weak. As seen in Figures \ref{fig:evolve}(e)-(h), the magnetic field shares
a similar structure to that of the velocity field, but it possesses a finer spatial scale as a result of the large
magnetic Prandtl number. Here again the connectedness of strong magnetic fields across the core is evident, though one
must take into account the switch in color in following a field concentration across the core center as measured by the
sense of its radial field.

\section{Effects of Magnetism on Core Dynamics} \label{sec:compare}

To illustrate the effective impact of the magnetic field upon the structure of the flow itself as well as upon the
resulting angular velocity distribution within the core, it is useful to compare the hydrodynamic case {\sl H4} to its
magnetic counterpart case {\sl M4}. To this end, one may first consider the differential rotation captured in Figure
\ref{fig:hdvmhd}, which displays the time-averaged and azimuthally averaged angular velocity achieved in those two
cases. It is striking that whereas the hydrodynamic case possesses very pronounced variations in angular velocity
$\Omega$ in the core with both radius and latitude, the magnetic companion case is nearly in uniform rotation. In case
{\sl H4} such variations represents up to 60\% of the mean angular velocity $\Omega_0$, but barely reach 10\% in case
{\sl M4}. As seen in Figures \ref{fig:hdvmhd}(a), (b), (d), and (e), both simulations exhibit columnar structures in
$\Omega$ within the convective cores resulting from angular momentum redistribution, leading to a central column of
strongly retrograde zonal flow in the hydrodynamic case, and one of weakly prograde flow in the magnetic case. Such
columnar retrograde states of angular velocity have already been realized within hydrodynamical 3D simulations of core
convection in rotating A-type stars by \citet{browning04} and \citet{cai11} and are the direct consequence of states
with low Rossby number.  Likewise, follow-on simulations of core convection dynamos in A-type stars by \citet{brun05}
showed that the presence of magnetism largely suppressed the differential rotation within the core.

\begin{figure}[t!]
  \begin{center}
    \includegraphics[width=0.45\textwidth]{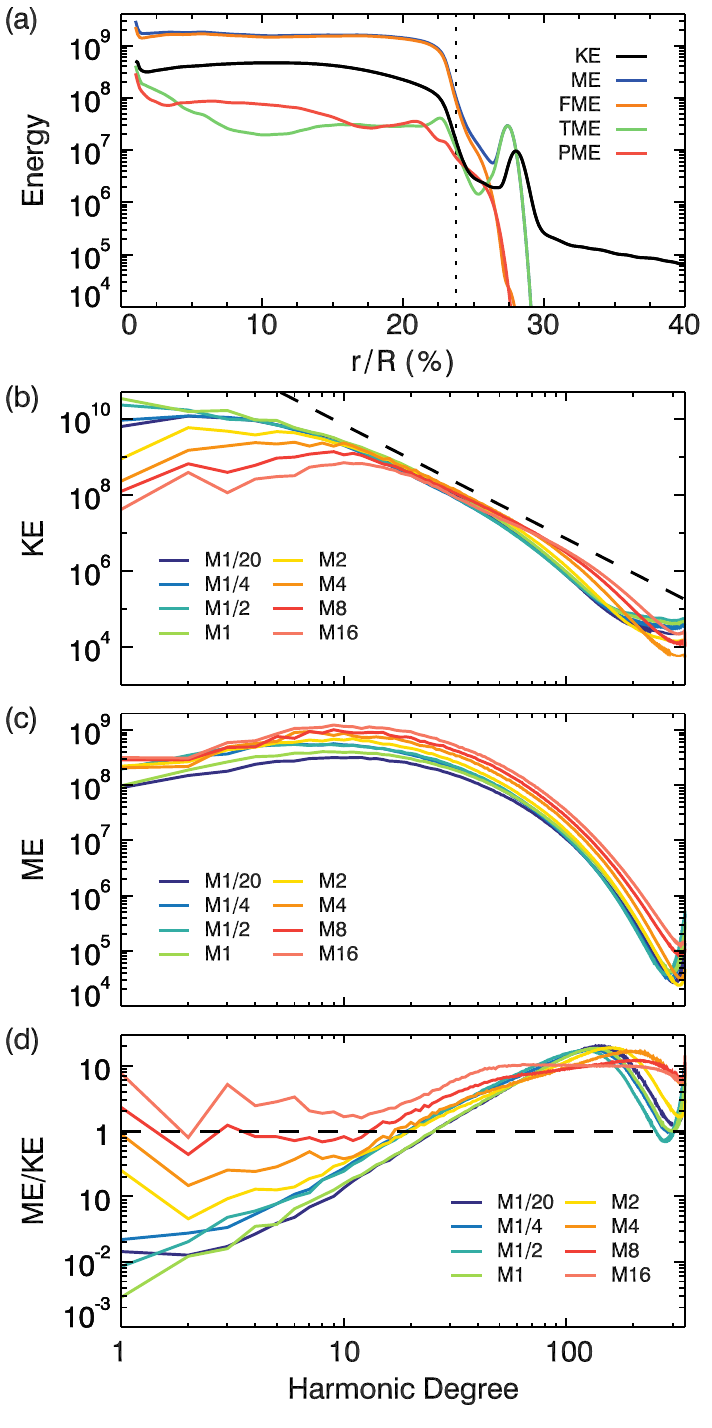}

    \figcaption{Radial distribution and spectra of magnetic and kinetic energies. (a) Temporally and longitudinally
      averaged radial energy distribution in case {\sl M16}, showing total kinetic energy (KE, black), total magnetic
      energy (ME, blue), fluctuating magnetic energy (FME, orange), mean toroidal magnetic energy (TME, green), and mean
      poloidal magnetic energy (PME, red).  The vertical dotted line indicates the location of the convective core
      boundary.  (b) Kinetic energy spectra as a function of spherical harmonic degree $\ell$, averaged in time and by
      quadratically summing the azimuthal components of order $m$. The dashed line provides a reference slope of
      $\ell^{-3}$. (c) Companion magnetic energy spectra with $\ell$.  (d) The ratio of the averaged kinetic and
      magnetic energy spectra. The dashed line illustrates the point of equipartition.  The spectra are taken at
      $r/R=15\%$. The temporal averages were performed over 134 days. \label{fig:spectra}}
  \end{center}
\end{figure}

To analyze how the angular momentum is continuously redistributed within the convective core to achieve such differing
states, it is convenient to appeal to the individual transport terms as presented in the appendix. Furthermore, it is
practical to focus on the cylindrical components of the angular momentum flux, given that the quantity of primary
interest is the radial differential rotation of the core. With such a focus, it becomes clear from Figures
\ref{fig:hdvmhd}(c) and (f) that there is a transfer from the influence of the viscous stresses in the hydrodynamic case
{\sl H4} to the Maxwell stresses in the MHD case {\sl M4}. In particular, the role of the viscous transport of angular
momentum is replaced by transport through the fluctuating Maxwell stress. Indeed, when both the fluid Reynolds number
and the Rossby number are beyond some critical values, and the magnetic Prandtl number is greater than unity, this
behavior should be expected. The reason for this is that, when temporally and vertically averaged in simulations with
such a parameter regime, the cylindrically outward angular momentum transport arising from the meridional flows (MC in
Figures \ref{fig:hdvmhd}(c) and (f)) should be minimal. In the hydrodynamic cases, this leaves a primary balance between
the Reynolds stress and the viscous stress, as is seen in Figure \ref{fig:hdvmhd}(c). Likewise, once magnetic angular
momentum transport mechanisms are available in a regime of high Prandtl number, the primary balance is between the
Reynolds stress and the fluctuating Maxwell stress.

Consider next the radial distribution of kinetic and magnetic energy in the core and overshooting region of case {\sl
  M16} to further elucidate the interplay between the flows and the dynamo-generated magnetic fields.  As seen in Figure
\ref{fig:spectra}(a), the radial dependences of kinetic and magnetic energies are found to be nearly constant throughout
the core and then to decrease by about two orders of magnitude in the overshooting region, which extends further by 6\%
in proportional radius beyond the convective core boundary. The magnetic field clearly contains more energy than the
flows in the convective core or those in the overshooting region. Given that the initial condition had zero magnetic
field outside the core, there is no magnetic energy in the radiative exterior, except for a slow diffusion there due to
the tapered radial profile of the magnetic diffusivity $\eta$ employed here. The peak in the kinetic and magnetic energy
near $r/R = 27\%$ corresponds to the depth at which the diffusivity coefficients rapidly drop to their minimum values in
the radiative exterior. It may arise due to the action of large-scale shear as the rotation profile converges toward a
solid body rotation of the radiative exterior, as indicated by TME being enhanced in that area.  The other cases show
similar radial distributions of energy, with it being nearly constant throughout the core and dropping rapidly outside
of it, and with the fluctuating magnetic energy (FME) being larger than the energy in the mean fields (TME $+$ PME).

Figures \ref{fig:spectra}(b) and (c) present for all magnetic cases the kinetic and magnetic energy spectra with spherical
harmonic degree $\ell$, averaged in time and quadratically summed over azimuthal order $m$, at a radius of $r/R=15\%$.
The kinetic energy spectra in Figure \ref{fig:spectra}(b) have three characteristic regions: the low-degree
(large-scale) range, the inertial range, and the high-degree (small-scale) dissipative range. The most striking feature is the
diminution in the amplitude of the large-scale flows as the rotation rate increases (going from {\sl M1/20} to {\sl
  M16}).  The inertial range at intermediate $\ell$-values, namely the region that is likely the result of a
self-similar cascade of kinetic energy, is fairly similar among all the cases. The largest differences between the cases
are where the transition points are between those three characteristic regions. The relative smoothness of KE at low
$\ell$ arises from the absence of axisymmetric even modes that are usually connected to the presence of strong mean
zonal flows representing differential rotation that has been wiped out in these magnetic cases.

\begin{table*}[t!]
   \begin{center}
   \begin{tabular}{cccccccccccc}
   \multicolumn{12}{c}{\bf Table 2} \\
   \multicolumn{12}{c}{Global Properties of the Evolved Core Convection} \\
   \hline
   \hline
   Case & $\mathrm{KE}/10^8$ & $\mathrm{CKE}/10^8$ & $\mathrm{DKE}/10^8$  & $\mathrm{MKE}/10^8$ & $\mathrm{ME}/10^8$ & 
   $\mathrm{FME}/10^8$ & $\mathrm{TME}/10^8$ & $\mathrm{PME}/10^8$ & $\mathrm{ME}/\mathrm{KE}$ & $B_{rms}$ & $B_{max}$ \vspace{0.025truein} \\
   \hline
   {\sl M1/20} & 23.3 & 19.8 (84.9\%) & 1.63  (7.0\%) & 1.894  (8.1\%) &  3.76 &  3.59 (95.4\%) & 0.10 (2.7\%) & 0.07 (1.9\%) & 0.16 & 0.10 & 1.46 \\
   {\sl M1/4}  & 25.1 & 18.6 (74.0\%) & 3.73 (14.9\%) & 2.806 (11.1\%) &  5.77 &  5.45 (94.5\%) & 0.19 (3.3\%) & 0.13 (2.2\%) & 0.23 & 0.12 & 1.33 \\
   {\sl M1/2}  & 31.7 & 22.9 (72.3\%) & 7.11 (22.4\%) & 1.687  (5.3\%) &  5.94 &  5.58 (93.9\%) & 0.26 (4.4\%) & 0.10 (1.7\%) & 0.19 & 0.12 & 1.29 \\
   {\sl M1}    & 29.0 & 27.0 (93.0\%) & 1.51  (5.2\%) & 0.535  (1.8\%) &  4.95 &  4.65 (94.0\%) & 0.23 (4.5\%) & 0.07 (1.5\%) & 0.17 & 0.11 & 1.43 \\
   {\sl M2}    & 12.6 & 11.1 (87.9\%) & 1.13 (10.7\%) & 0.182  (1.4\%) &  7.27 &  6.78 (93.3\%) & 0.39 (5.3\%) & 0.10 (1.4\%) & 0.58 & 0.13 & 1.20 \\
   {\sl H4}    & 97.5 & 15.5 (15.9\%) & 81.8 (83.8\%) & 0.280  (0.3\%) &  --   &  --            &  --          & --           & --   & --   & --   \\
   {\sl M4}    & 8.01 & 7.64 (95.4\%) & 0.33  (4.1\%) & 0.042  (0.5\%) &  9.39 &  8.77 (93.4\%) & 0.48 (5.1\%) & 0.14 (1.5\%) & 1.17 & 0.15 & 1.25 \\
   {\sl M8}    & 4.65 & 4.42 (95.2\%) & 0.21  (4.5\%) & 0.015  (0.3\%) & 11.20 & 10.53 (94.0\%) & 0.35 (3.2\%) & 0.31 (2.8\%) & 2.41 & 0.17 & 1.58 \\
   {\sl M16}   & 2.78 & 2.63 (94.3\%) & 0.15  (5.5\%) & 0.006  (0.2\%) & 13.96 & 13.24 (94.9\%) & 0.34 (2.4\%) & 0.38 (2.7\%) & 5.02 & 0.19 & 2.11 \\
   \hline
   \end{tabular}
   \end{center}
   \tablecomments{Total kinetic energy density (KE) and those of the convection (CKE), differential
     rotation (DKE), and meridional circulation (MKE). Total magnetic energy density (ME) and of
     fluctuating (FME), toroidal (TME), and poloidal (PME). These energies are averaged in time and
     in the volume of the core and are shown in units of $\mathrm{erg}\, \mathrm{cm}^{-3}$, and
     with component percentages enclosed parenthetically. The rms magnetic field strength
     $B_\mathrm{rms}$ and peak magnetic field strengths $B_\mathrm{max}$ are shown in units of
     MG.}
\end{table*}

The magnetic energy spectra in Figure \ref{fig:spectra}(c) show gradual change with rotation rate being mostly
self-similar.  The amplitude of the magnetic energy increases monotonically with rotation rate or decreasing Rossby
number.  The scales containing the most magnetic energy are moderate, with the broad peak in the spectrum occurring
around $\ell=12$ rather than at the lowest wavenumbers. Such an energy distribution could be termed an
intermediate-scale dynamo, i.e., the largest-scale modes are not the dominant ones but rather that the scales within the
inertial range contribute the most.  Figure \ref{fig:spectra}(d) displays the ratio of the magnetic to the kinetic
energy as a function of harmonic degree at the same depth as panels (b) and (c). This shows the super-equipartition (or
sub-equipartition) state of a given simulation for all resolved scales. What is striking is the change of the ME/KE
ratio at low $\ell$ degree by almost three orders of magnitude as one goes from case {\sl M1/20} to case {\sl M16}.  In
particular, ME for cases {\sl M8} and {\sl M16} is now equal to or larger than KE at all scales.  This change in
behavior at low $\ell$ degree between slowly and rapidly rotating cases may be attributable to an increasing influence
of the Coriolis force on the largest scales, as will be further discussed in \S 6.  When comparing Figures
\ref{fig:spectra}(b) and \ref{fig:spectra}(c) and considering the spectra as the rotation rate increases, it is clear
that the increase in the ME/KE ratio at intermediate and smaller scales is primarily due to changes in the kinetic
energy spectrum. Moreover, it is evident that all cases exhibit a ME/KE ratio that is greater than unity for scales with
$\ell\ge 10$ for case {\sl M8} and $\ell\ge 20$ for cases {\sl M1/20} and {\sl M1}. This characteristic is another
hallmark of the classical ``small-scale'' dynamo, which in turn is expected for magnetic Prandtl numbers greater than a
few. The ME/KE ratio grows as large as 20 for scales with $\ell\approx 130$ in case {\sl M1/20} and $\ell\approx 230$ in
case {\sl M16}. This occurs because the viscous diffusion is large enough to begin to quench the flows at those scales
before the magnetic field enters its dissipative range of scales, which are smaller given the large magnetic Prandtl
number \citep{scheko04}, and this is a realistic situation in such B-type stars.

\section{Examining a Convective Core Dynamo} \label{sec:examine}

Some of the salient features of these convective core dynamo simulations are captured in Table 2. In particular, it is
worth highlighting the ratio of the magnetic energy to the kinetic energy, as well as the peak and rms magnetic field
strengths in the core. The energy densities provided in that table are defined as follows:

\begin{align}
  & \mathrm{DKE} = \frac{1}{2}\orho\avg{\vcp}^2,  \; \mathrm{MKE} = \frac{1}{2}\orho\left(\avg{\vcr}^2 + \avg{\vct}^2\right), \nonumber \\
  &\mathrm{CKE} = \frac{1}{2}\orho\left(\vv-\avg{\vv}\right)^2, \; \mathrm{TME} = \frac{1}{8\pi}\avg{\Bcp}^2, \nonumber \\ 
  &\mathrm{PME} = \frac{1}{8\pi}\left(\avg{\Bcr}^2 + \avg{\Bct}^2\right), \; \mathrm{FME} = \frac{1}{8\pi}\left(\vB-\avg{\vB}\right)^2, \nonumber\\
  &\mathrm{PE} = \frac{1}{8\pi}\left[\curl{\curl{\left(C\rht\right)}}\right]^2, \; \mathrm{TE} = \frac{1}{8\pi}\left[\curl{\left(A\rht\right)}\right]^2,
\end{align}

\noindent with the $\langle\rangle$ denoting an average in longitude, $\vv'=\vv-\avg{\vv}$ the fluctuating velocity,
$\vB'=\vB-\avg{\vB}$ the fluctuating magnetic field, and $\avg{\vv}$ and $\avg{\vB}$ the axisymmetric velocity and
magnetic field respectively. As part of a standard solenoidal decomposition of the magnetic field, the poloidal magnetic
potential is defined as $C$ and the toroidal magnetic potential as $A$. The total kinetic energy density KE is
$\mathrm{DKE}+\mathrm{MKE}+\mathrm{CKE}$ and the total magnetic energy density ME is
$\mathrm{TME}+\mathrm{PME}+\mathrm{FME}$.  These values are averaged in time and within the volume of the core. For the
choices of parameters here, it is clear that both rms magnetic field and the peak magnetic fields increase with faster
rotation.  Moreover, the ratio of magnetic to kinetic energy increases as the rotational constraint on the flows
increases. Additionally, Table 2 shows a decreasing magnetic energy in the mean magnetic fields (PME, TME) as a
percentage of the total magnetic energy. The mean magnetic fields are weakest at the slowest and fastest rotation rates,
with their fractional content rising slightly at moderate rotation rates.  Moreover, the decreasing kinetic energy in
each more rapidly rotating case serves to further enhance the ratio of the magnetic to kinetic energy as the magnetic
energy grows more slowly than the rate of decrease of the kinetic energy with increasing rotation rate. Indeed it is
fascinating that the dynamo sustains super-equipartition magnetic fields once above a certain rotational threshold. This
feature will be the primary focus of the subsequent sections.

\begin{figure}[t!]
  \begin{center}
    \includegraphics[width=0.45\textwidth]{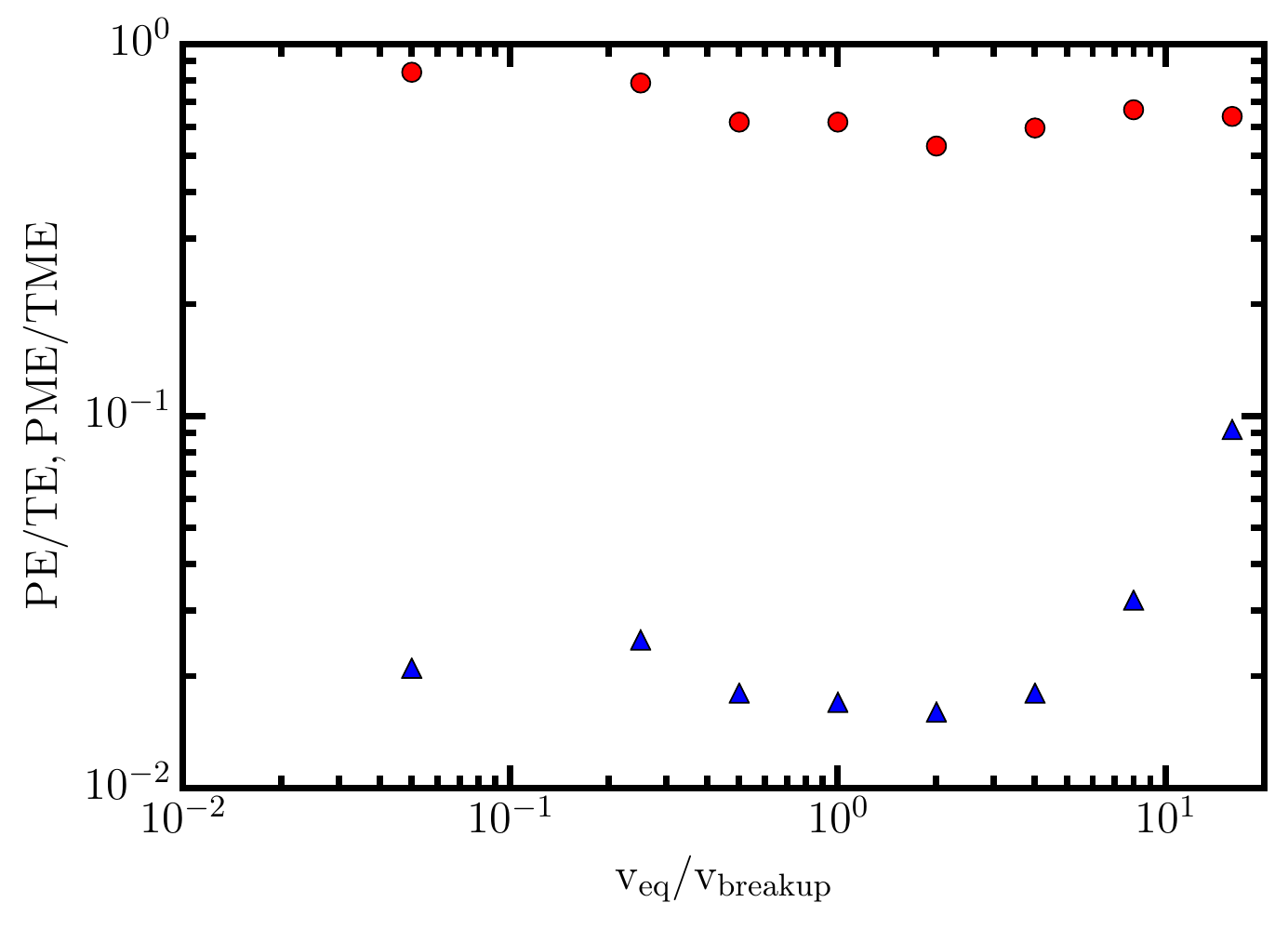}

    \figcaption{The ratio of total poloidal and toroidal magnetic energy (PE/TE) is shown as a red circle for each case,
      and the ratio of the mean poloidal and mean toroidal magnetic energy (PME/TME) is shown as a blue triangle for
      each case.\label{fig:energyratio}}
  \end{center}
\end{figure}

\begin{figure*}[t!]
  \begin{center}
    \includegraphics[width=\textwidth]{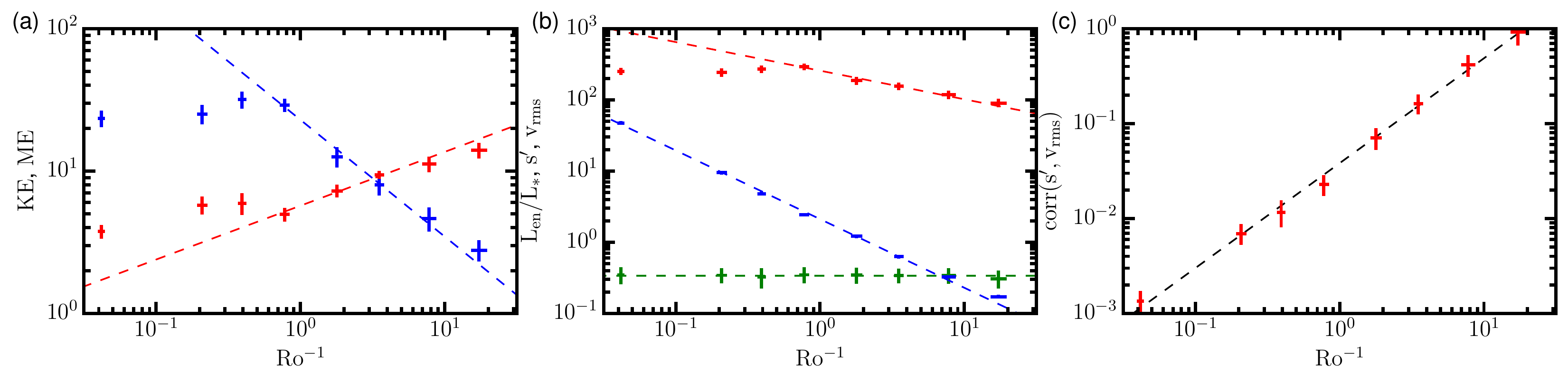}

    \figcaption{Scaling of kinetic and magnetic energy as well as components of the heat flux. (a) Kinetic (blue) and
      magnetic (red) energies varying with inverse Rossby number, in units of $10^8$~erg. (b) Peak of the shell-averaged
      enthalpy flux ($L_{en}/L_{*}$, green), rms entropy perturbation ($s'$, blue), and rms convective velocity
      ($\mathrm{v}_{rms}$, red) varying with inverse Rossby number. (c) The correlation of $s'$ and $\mathrm{v}_{rms}$,
      defined as $L_{en}/(s' \mathrm{v}_{rms} L_{*})$, normalized by this value for case {\sl M1/20}.  The standard
      deviation of the Rossby number and a given dependent quantity arising from its time variation are indicated by the
      length of each side of a data point's cross. The dashed lines are the best fits to the portions of the data that
      appear to have linear trends in log-log space. \label{fig:convscaling}}
  \end{center}
\end{figure*}

A simple theoretical scaling for the ratio of magnetic energy to kinetic energy was described in \S\ref{sec:scaling},
and shown in Figure \ref{fig:scaling}. This scaling relationship appears to describe the trend of the measurement of
$\mathrm{ME/KE}$ from the simulations quite well, as also shown in Figure \ref{fig:scaling}.  This is especially true of
the more rapidly rotating cases that fall along the magnetostrophic asymptote denoted with the dashed green line. This
asymptotic regime is an indication that the primary force balance in these cases is between the Lorentz and Coriolis
forces.  Considering the values of the dynamic Elsasser number ($\Lambda_D$) in Table 2, it is evident that this global
measure of the ratio of those forces is nearly unity, indicating their approximate equivalence.  In contrast, the cases
with a relatively large Rossby number tend to fall along a branch of nearly constant magnetic energy relative to the
kinetic energy, with that value being approximately 0.18.  These cases are dominated by a balance between the fluid
inertial forces and the Lorentz force. An exception to this relatively good agreement arises when the Rossby number of
the system reaches the critical threshold of $\mathrm{Ro}\approx 1$, which occurs roughly between cases {\sl M1} and
{\sl M2}.  In those cases, there appears to be a change in the topology of the fluctuating magnetic field, with it
becoming more dominantly toroidal. This transitional state can be seen in Figure \ref{fig:energyratio}, where there is a
minimum in the ratio of the total poloidal magnetic energy relative to the total toroidal magnetic energy near the
equatorial velocity consistent with case {\sl M2}.

The degree of equipartition of a given case is intimately tied to the independent scaling of the kinetic and magnetic
energies. These scalings are captured in Figure \ref{fig:convscaling}(a), where the data and their time-dependent
uncertainty are plotted along with the linear best fits to them. Shown in this manner, there is a clear transition from
a roughly constant scaling for both the magnetic and kinetic energies to a power-law scaling (linear trend in this
log-log plot) near the point where the Rossby number becomes unity for case {\sl M1}. At that point, the kinetic energy
begins to plunge and the magnetic energy grows slowly, so their ratio gives rise to the power-law scaling with inverse
Rossby number.  In particular, the kinetic energy scales as $\mathrm{KE}\propto \mathrm{Ro}^{0.82}$ and the magnetic
energy as $\mathrm{ME}\propto \mathrm{Ro}^{-0.38}$, which is consistent with the scaling derived in \S\ref{sec:scaling}
(e.g., $\mathrm{ME/KE} \propto \mathrm{Ro}^{-1}$) to within the bounds of the measured uncertainty.

In Figure \ref{fig:convscaling}(b) the rms velocity, average entropy perturbation, and average maximum in enthalpy flux
are shown for each of the cases. As was shown in Figure \ref{fig:energy}, the growth of the enthalpy flux follows the
increase in the total energy flux arising from nuclear fusion processes deep in the core. It reaches a maximum value
when the radiative flux has increased sufficiently to carry the majority of the energy flux. That value and its
uncertainty are plotted.  From the figure, it is clear that this maximum value is consistent with being constant with
Rossby number. Though it is not shown for each case, the form of the shell- and time-averaged enthalpy flux remains
nearly constant as well. So, in order to account for the constant enthalpy flux and the decreasing kinetic energy, and
likewise for the decreasing rms velocity, the entropy perturbations must be assessed. In doing so, it becomes evident
that, contrary to what one might expect from mixing-length theory, the amplitude of the entropy perturbations also
decreases and indeed more quickly so than the rms velocity. In particular, they scale as
$\mathrm{v}_{rms}\propto\mathrm{Ro}^{0.40}$ and $s'\propto\mathrm{Ro}^{0.96}$.

Since the radial enthalpy flux $L_{en} \propto \avg{s' \mathrm{v}_r}$, the only remaining possibility is that the
correlation between the two fields increases rapidly with decreasing Rossby number. This correlation is characterized by
$\mathrm{corr}(s',\mathrm{v}_r) \approx L_{en}/(L_* s' \mathrm{v}_{rms})$ and that value and its uncertainty are
illustrated in Figure \ref{fig:convscaling}(c) for each case, and it scales as
$\mathrm{corr}(s',\mathrm{v}_r) \propto \mathrm{Ro}^{-1.11}$.  So, whereas the amplitudes of velocity and entropy
decrease with decreasing Rossby number, their correlation increases rapidly and thereby stems any decrease in the
enthalpy flux.  Strikingly, their correlation increases by a factor of about 800. This occurs because coherent
structures form in a rapidly rotating system, yielding quasi-two-dimensional structures that can be very efficient at
heat transport \citep[e.g.,][]{yadav16}. Indeed, it is well known that structure formation is typical in rapidly
rotating convective systems, where, even in very turbulent systems, convective structures that are aligned with the
rotation axis are formed that largely obey the spirit of the Taylor-Proudman constraint.  And, as will be illustrated in
Figures \ref{fig:overlap}(a)-(c), such coherent vortical structures also prove to be efficient at building
super-equipartition magnetic fields, given their ease in enhancing the local kinetic helicity.

\section{Generating Global-Scale Magnetism} \label{sec:genmag}

\begin{table}[t!]
   \begin{center}
   \begin{tabular}{ccccccccc}
   \multicolumn{9}{c}{\bf Table 3} \\
   \multicolumn{9}{c}{Balance of Magnetic Energy Generation Mechanisms} \\
   \hline
   \hline
   Case & $\FFI$ & $\FMV$ & $\FMB$ & $\PFI$ & $\PMI$ & $\TFI$ & $\TMI$ & $\TRD$ \vspace{0.025truein} \\
   \hline
   {\sl M1/20} & 87\% & 10\% & 3\% & 68\% & 32\% &  79\% &  21\% & -100\% \\
   {\sl M1/4}   & 78\% & 20\% & 2\% & 41\% & 59\% &  48\% &  52\% & -100\% \\
   {\sl M1/2}   & 77\% & 22\% & 1\% & 47\% & 53\% &  42\% &  58\% & -100\% \\
   {\sl M1}       & 89\% &  6\% & 5\% & 75\% & 25\% &  98\% &   2\% & -100\% \\
   {\sl M2}       & 88\% &  8\% & 5\% & 87\% & 13\% &  70\% &  30\% & -100\% \\
   {\sl M4}       & 94\% &  2\% & 5\% & 93\% &  7\%  &  96\% &   4\% & -100\% \\
   {\sl M8}       & 95\% &  2\% & 3\% & 74\% & 26\% & 100\% &  -9\% &  -91\% \\
   {\sl M16}     & 95\% &  1\% & 4\% & 71\% & 29\% & 100\% &  -6\% &  -94\% \\
   \hline
   \end{tabular}
   \end{center}

   \tablecomments{The time-averaged and core-volume-integrated magnetic energy generation rates for
     the mechanisms shown in Equations (\ref{eqn:evofme})-(\ref{eqn:evotme}).
     Positive values indicate generation and negative values dissipation of magnetic energy. 
     The values for $\FRD$ and $\PRD$ are all $-100\%$, and so are not shown explicitly.}
\end{table}

\begin{figure*}[t!]
  \begin{center}
    \includegraphics[width=0.9\textwidth]{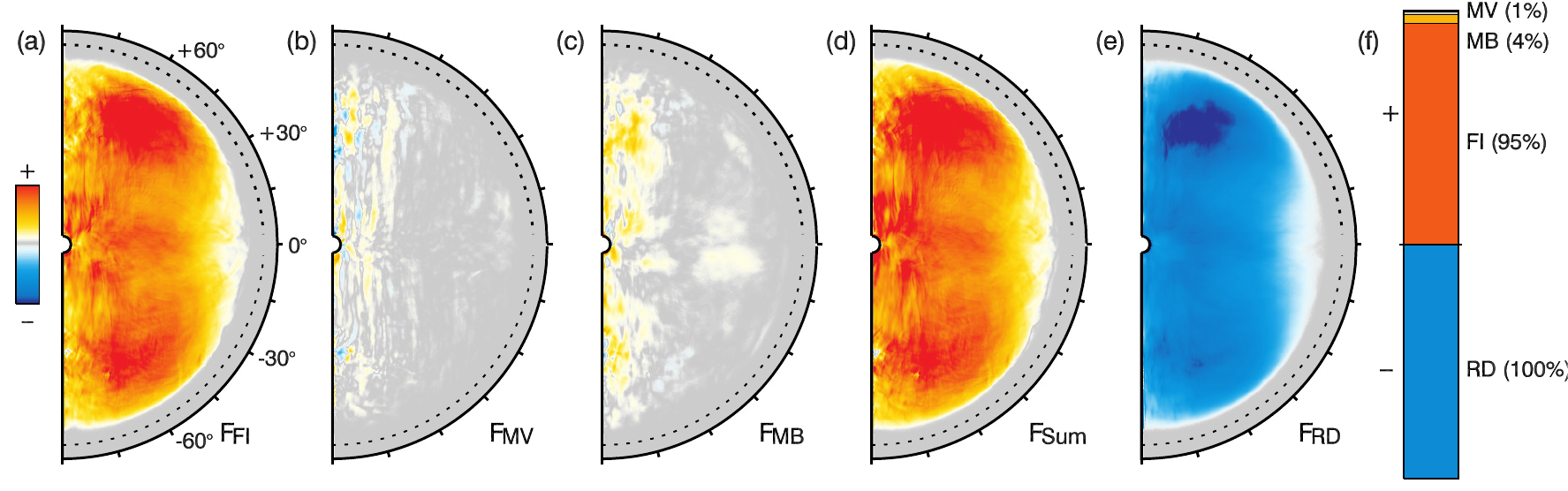}

    \figcaption{Generation of fluctuating magnetic energy in case {\sl M16}. The time-averaged and azimuthally averaged
      terms in Equation (\ref{eqn:evofme}) reveal that with a sufficiently long average a balance between those terms is
      achieved, with the terms shown as (a) fluctuating induction ($\FFI$), (b) induction from the axisymmetric velocity
      field acting on the fluctuating magnetic field ($\FMV$), (c) induction from the fluctuating velocity acting on the
      axisymmetric magnetic field ($\FMB$), (d) sum of the production terms (a)-(c), and (e) dissipation from resistive
      diffusion.  (f) The balance of the terms when integrated over the volume of the core. \label{fig:fmegen}}
  \end{center}
\end{figure*}

There are three components of the magnetic energy generation that must be considered to see the complete picture of how
the magnetic fields are sustained. These components of the total magnetic energy are the production (or destruction) of
fluctuating (FME), poloidal (PME), and toroidal magnetic energy (TME). Averaging the induction equation in longitude
(Equation (\ref{eqn:ashind})) yields the following evolution equations for the components of the total magnetic energy:

\begin{align}
  \displaystyle \ddtime{\FME} &= \overbrace{\bigavg{\frac{\vB'}{8\pi}\boldsymbol{\cdot}\curl{\left(\vv'\cross\vB'\right)}}}^{\FFI}
  + \overbrace{\bigavg{\frac{\vB'}{8\pi}\boldsymbol{\cdot}\curl{\left(\avg{\vv}\cross\vB'\right)}}}^{\FMV} \nonumber \\
  & + \overbrace{\bigavg{\frac{\vB'}{8\pi}\boldsymbol{\cdot}\curl{\left(\vv'\cross\avg{\vB}\right)}}}^{\FMB} 
  - \overbrace{\bigavg{\frac{\vB'}{8\pi}\boldsymbol{\cdot}\curl{\left(\eta\curl{\vB'}\right)}}}^{\FRD}, \label{eqn:evofme}
\end{align}

\begin{align}
  \displaystyle \ddtime{\PME} &= \overbrace{\frac{\MPF}{8\pi}\boldsymbol{\cdot}\curl{\left(\avg{\vv}\cross\avg{\vB}\right)}}^{\PMI}
  + \overbrace{\frac{\MPF}{8\pi}\boldsymbol{\cdot}\curl{\avg{\vv'\cross\vB'}}}^{\PFI} \nonumber \\
  & - \overbrace{\frac{\MPF}{8\pi}\boldsymbol{\cdot}\curl{\left(\eta\curl{\avg{\vB}}\right)}}^{\PRD}, \label{eqn:evopme}
\end{align}

\begin{align}
  \displaystyle \ddtime{\TME} &= \overbrace{\frac{\MTF}{8\pi}\pht\boldsymbol{\cdot}\curl{\left(\avg{\vv}\cross\avg{\vB}\right)}}^{\TMI}
  + \overbrace{\frac{\MTF}{8\pi}\pht\boldsymbol{\cdot}\curl{\avg{\vv'\cross\vB'}}}^{\TFI} \nonumber \\
  & - \overbrace{\frac{\MTF}{8\pi}\pht\boldsymbol{\cdot}\curl{\left(\eta\curl{\avg{\vB}}\right)}}^{\TRD}. \label{eqn:evotme}
\end{align}

Here $\mathrm{F}$ indicates a fluctuating magnetic energy generation term, $\mathrm{P}$ a mean poloidal energy
generation term, and $\mathrm{T}$ a mean toroidal energy generation term. The subscripts of these monikers denote
production in turn by mean induction ($\mathrm{MI}$), fluctuating induction ($\mathrm{FI}$), and resistive diffusion
($\mathrm{RD}$). For the fluctuating magnetic energy, there is induction arising from the action of the mean velocity on
the fluctuating magnetic field ($\mathrm{MV}$), and induction arising from the action of the fluctuating velocity on the
mean magnetic field ($\mathrm{MB}$).

The values for each of the terms in Equations (\ref{eqn:evofme})-(\ref{eqn:evotme}) averaged over time and over the core
volume are given in Table 3.  Clearly, when also considering the amount of energy contained in the various components of
the magnetic field provided in Table 2, the primary mechanism for generating magnetism in these simulations is through
fluctuating inductive processes. Namely, it is generated through those processes that are composed of the product of
fluctuating velocity and magnetic fields with either the mean field ($\PFI$ or $\TFI$) or the fluctuating magnetic field
itself ($\FFI$). Surprisingly, the induction from the mean fields can act to dissipate magnetic energy in the most
rapidly rotating cases. However, the vast majority of the dissipation of magnetic energy is through resistive diffusion,
balancing its generation by the fluctuating inductive processes.

As an example, and to exhibit the radial and latitudinal distribution of the energy generation or dissipation, each of
the terms in Equation (\ref{eqn:evofme}) is shown for case {\sl M16} in Figure \ref{fig:fmegen}. Simply comparing Figure
\ref{fig:fmegen}(a) and \ref{fig:fmegen}(e) shows that the dominant balance is between the fluctuating induction and
resistive dissipation, with the mean fields playing very minor but supportive roles in the generation of fluctuating
magnetic energy. One important feature is the nearly uniform distribution of magnetic energy production through the
fluctuating induction process, which in turn induces the uniform distribution of the magnetic energy
dissipation. Indeed, these two features are shared in all the cases, with only small differences in the spatial
distribution of where the fluctuating induction takes place.

Turning to the distribution of the values averaged over time and over the core volume for the components of the total
magnetic energy provided in Table 2, it is clear that more than 93\% of the magnetic energy resides in the fluctuating
magnetic energy for all the cases. Thus, given the small amount of energy contained in the mean magnetic fields, they
play a small role in these dynamos.  Furthermore, as seen in Table 2, the energy contained in the differential rotation
is in most cases also small compared to the energy contained in the fluctuating velocity field. Therefore, since both
those mean fields have little energy, the dynamo operating in the cases presented here cannot be characterized as the
$\alpha-\Omega$ dynamos that appear to operate so keenly in the less massive main-sequence stars with an exterior
convective envelope. It is difficult to classify these core dynamos in terms of a classical mean-field dynamo at
all. The reasoning is simple: while the mean-field component of the magnetic field may satisfy some mean-field theoretic
classification, there is no such analogy for a system where the fluctuating magnetic energy is dominant. However, one
may construct a mean-field model for the evolution of the azimuthally averaged fluctuating magnetic energy, as it is
quadratic in the field and does not vanish under the averaging operator. One such model, based upon the $\alpha$-effect
mean-field theoretic paradigm, is the fairly general bilinear form

\begin{align}
  \frac{\partial \avg{\vB'^2}}{\partial t} &\approx \avg{B'_i\epsilon_{ijk}\partial_j\alpha_{kl}B'_l}
        - \avg{B'_i\epsilon_{ijk}\partial_j\beta_{klm}\partial_l B'_m},
\end{align}

\noindent where $\alpha_{ij}$ is a component of a parameterization of the kinetic helicity that induces magnetic field
production. In its simplest form, under the assumption of isotropic $\alpha$ and $\beta$ tensors, this reduces to

\begin{align}
  \frac{\partial \avg{\vB'^2}}{\partial t} &\approx \alpha\avg{H_J} - \beta \avg{\vB'\cdot\curl{\vJ'}},
\end{align}

\noindent where $H_J$ is the current helicity. Evaluating this balance in a global average for each of these terms in
all of the magnetic cases (as detailed in Table 3) does confirm that the dynamo operating within these simulations is
dominated by the fluctuating generation terms.  When these magnetic energy generation balances are examined in an
azimuthal average for each case, they appear nearly identical to those balances seen for case {\sl M16} in Figure
\ref{fig:fmegen}, where there is both a local and a global balance between ohmic dissipation and magnetic energy
generation by fluctuating induction.  Hence, for all of these simulations, this balance is not only a global property
but it is also a local one.

\section{Lorentz Force Feedbacks} \label{sec:feedback}

Having seen how these dynamos are able to build magnetic field, the remaining question is how the flows can survive the
strengthening Lorentz forces associated with an increasing rotation rate, which is particularly relevant for the
super-equipartition cases. Indeed, the first quantity of note is the average magnetic energy contained in the core,
which is shown in Table 2.  It increases with rotation rate and soon exceeds the energy contained in the flows.  Since
the dominant component of the magnetic energy is the fluctuating magnetic field, the increasing magnetic field strength
cannot be associated with an increasingly strong mean field as might be expected if the primary source for the energy
for the field arose from a differential rotation. Instead, since the kinetic energy and the flow velocity are diminished
as the rotation rate increases, the only possible explanation for the continuing growth of the magnetic energy is that
the structure of the flows is such that they become inherently more efficient at building magnetic field. The same was
seen in Sections \ref{sec:examine} and \ref{sec:genmag} for the maintenance of the enthalpy flux and the production of
magnetic field.

\begin{figure}[t!]
  \begin{center}
    \includegraphics[width=0.45\textwidth]{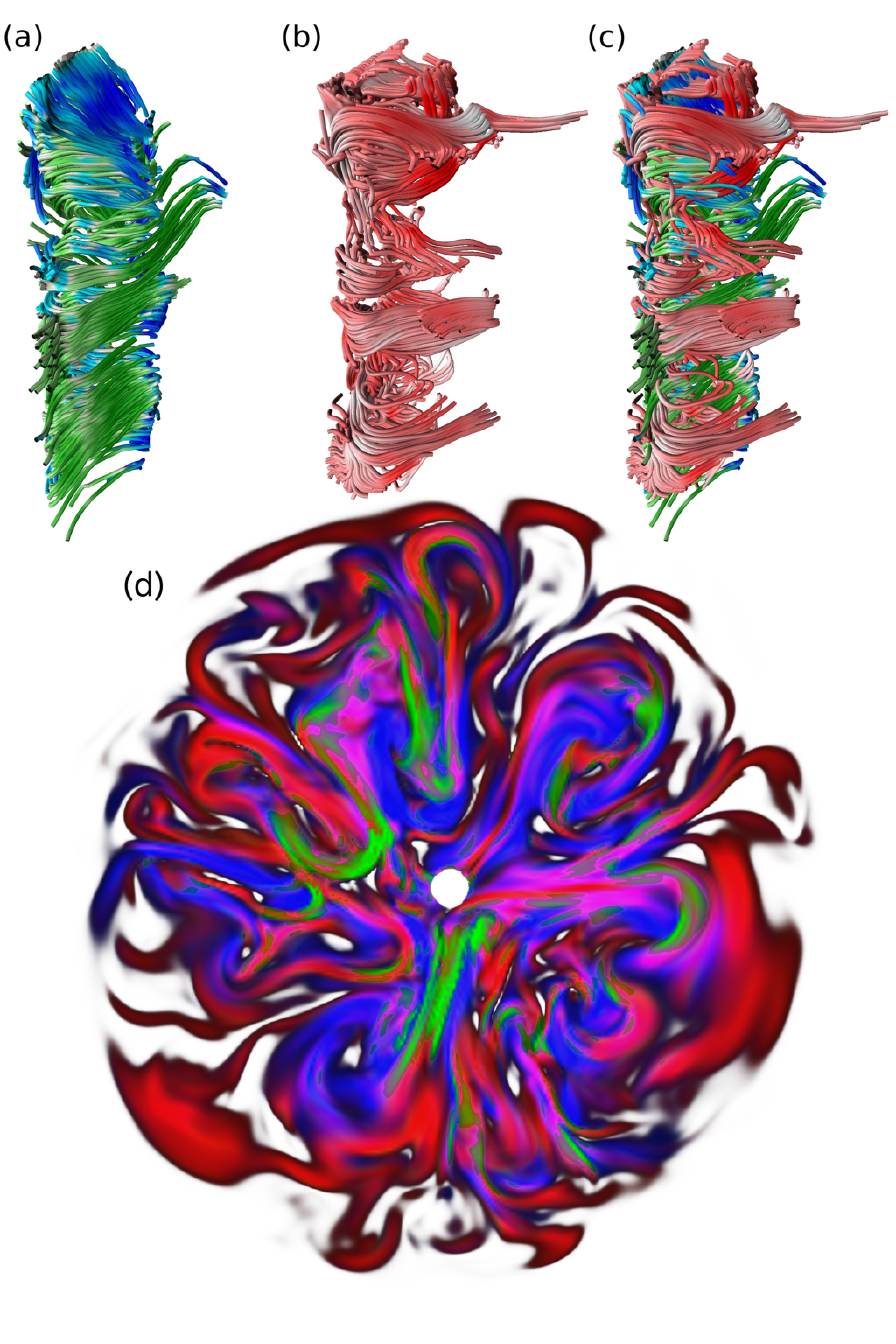}

    \figcaption{Focused visualization of a single convective column and the magnetic field associated with it and of the
      displacement of velocity and magnetic structures in case {\sl M16}. (a) The flow field of the convective column,
      with the color given by the value of the vertical velocity $\mathrm{v}_z$ (negative values are green, positive
      values are blue). (b) The magnetic field of the convective column, with the color given by the value of the
      vertical magnetic field $B_z$ (negative values are white, positive values are red). (c) The combined flow fields,
      colors as before. (d) The relative positions of magnetic and kinetic energy structures in a snapshot. Regions of
      large magnetic energy are in red, large kinetic energy in blue, and regions of overlap are purple. Green regions
      indicate where the Lorentz work is large. \label{fig:overlap}}
  \end{center}
\end{figure}

Structure formation and increased field correlation do not explain how the convective flows are able to avoid being
quenched by the strong magnetic fields induced at faster rotation rates. To see how those flows and magnetic fields
avoid mutual destruction, consider Figure \ref{fig:overlap}. This figure demonstrates features that likely lead to the
avoidance of disruptively strong Lorentz forces in super-equipartition cases. Figures \ref{fig:overlap}(a)-(c) show the
velocity field, magnetic field, and how they are entangled for a single convective column in case {\sl M16}. The first
aspect to note is that the velocity field is volume-filling, whereas the magnetic field is much more fibril and also
forms sheets rather than a more solid composite structure. When these two structures are allowed to visually intersect,
parts of the two overlap. However, much of the magnetic field is spatially adjacent to the velocity field structure,
effectively forming a magnetic sheath around the convective column. Indeed, such increasingly filamentary magnetic field
structures are somewhat expected for high-$\mathrm{Pm}$ systems \citep{scheko04}, which are typical of such B-type
stars.

Such spatial displacements of magnetic and flow structures can also be seen in Figure \ref{fig:overlap}(d), where the
kinetic energy, magnetic energy, and Lorentz work are co-rendered in a vertical average of a horizontal slab whose center
is located at the equator. The magnetic energy is rendered in red tones, the kinetic energy in blue tones, and the
Lorentz work in green tones. The regions of overlapping kinetic and magnetic energy structures are shown in purple. This
makes clear that the volume-filling fraction of regions where the magnetic and kinetic energy structures are co-spatial
is relatively small.  Likewise, the regions where the Lorentz force is doing work on the flows are fairly small relative
to the total volume, further reflecting the system's action to reduce the volume where the magnetic and velocity
structures overlap.

\begin{figure}[t!]
  \begin{center}
    \includegraphics[width=0.45\textwidth]{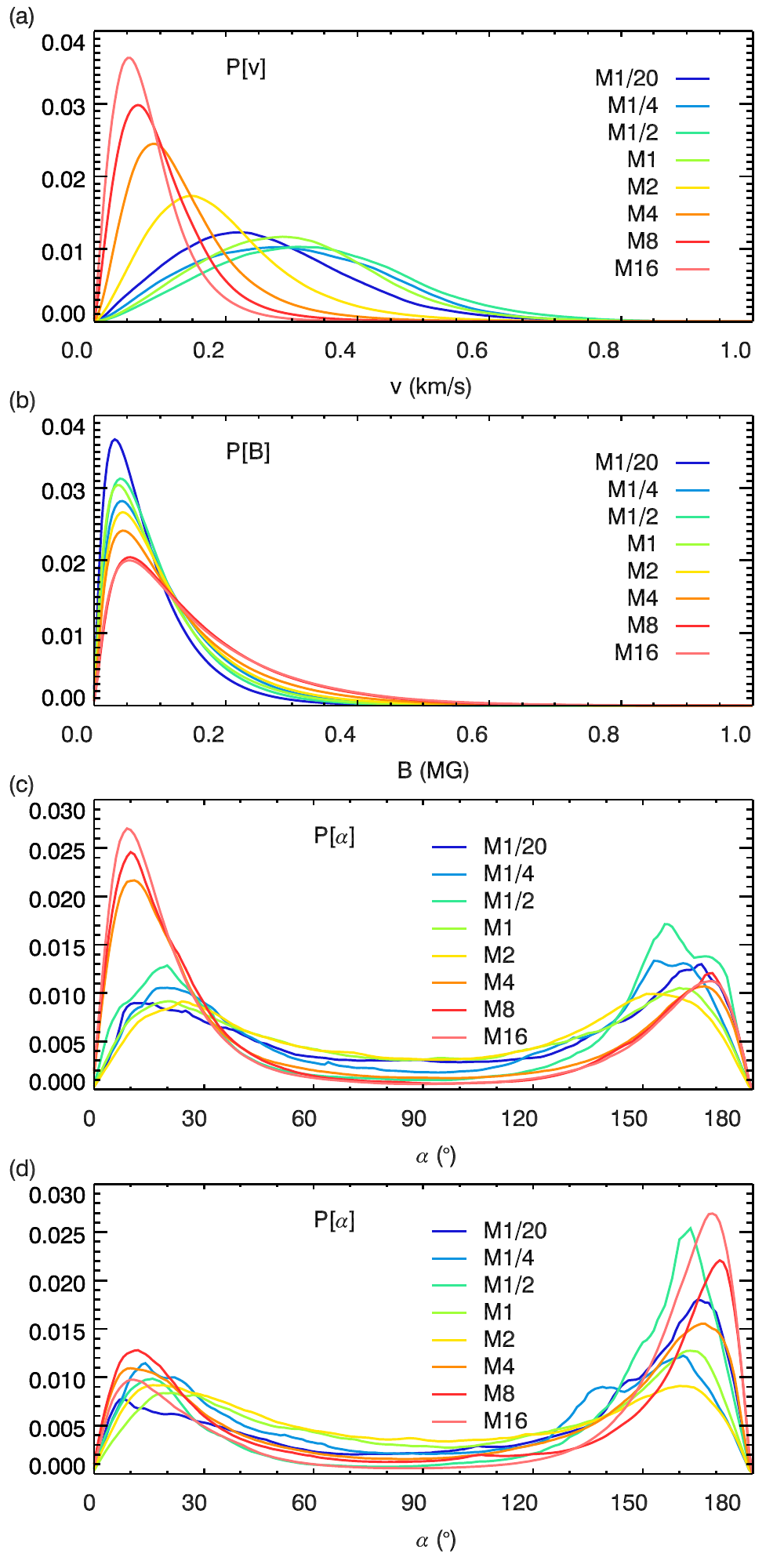}
    
    \figcaption{Probability distribution functions (PDFs) for the MHD simulations. Linearly scaled plots of (a) velocity
      magnitude and (b) magnetic field magnitude PDFs. Distribution of the relative angle between the magnetic field and
      the velocity, with the PDF shown for (c) the northern hemisphere and (d) the southern hemisphere. The chosen level
      set is the 90\% level of the joint cumulative distribution function for (c) and (d). The sampled points are thus
      restricted to the set intersection of regions of increasingly large kinetic and magnetic energy and thus show
      the increasing alignment of the fastest flows and strongest magnetic fields. \label{fig:pdfs}}
  \end{center}
\end{figure}

Indeed, even in those regions where the two fields overlap, the arrangement of the magnetic and velocity fields is such
that it reduces both the magnetic field generation and the Lorentz force.  The suppression of magnetic induction occurs
because of the increasing alignment of strong flows and fields, as is illustrated in Figures \ref{fig:pdfs}(c) and
(d). There the probability distribution function (PDF) of the angle between the magnetic field and the velocity field is
shown for fields that are above the strength of $90\%$ of the cumulative distribution function (CDF), which are flows of
about $\mpers{200}$ and magnetic fields about $350~\mathrm{kG}$. It is of further note that the magnetic field is
aligned with the velocity field in the northern hemisphere, whereas it is anti-aligned in the southern hemisphere.  For
the full PDF of this angle, it too is bimodal, but is less extreme and is north-south symmetric.  Although it is not
shown explicitly because the PDF is effectively a simple Lorentzian for the angle between the velocity field and the
Lorentz force, those PDFs are centered at $\dgr{90}$. This implies that these convective dynamos arrange themselves such
that the Lorentz work is minimized for all the level sets along the joint CDF of the magnetic and kinetic energy.

Considering the auto- and cross-correlations of the magnetic and velocity fields in space and time, the degree of
overlap or lack thereof can be quantified.  Figures \ref{fig:pdfs}(a) and (b) reveal that there are significant
influences on the nature of the velocity field and magnetic field as a function of rotation.  In particular, the median
value of the velocity decreases by about a factor of three, whereas the tails of the velocity distribution become more
truncated. This gives rise to a relatively leptokurtic PDF at faster rates of rotation.  The PDF of the magnetic field
amplitude, on the other hand, is quite similar in shape for each case.  Primarily, it only decreases in amplitude
slightly, which is simply a reflection of the longer tails (increased kurtosis) of the distributions at faster rotation
rates.

In analogy to hydrodynamic simulations that show a transition from prograde to retrograde equatorial differential
rotation that depends upon the Rossby number \citep{gastine14}, the transition from a sub-equipartition to a
super-equipartition state seen in \S\ref{sec:scaling} for these convective core dynamos is realized when the average
Rossby number of the convective flows drops below unity. In the regime of low Rossby number, the dynamo can sustain a
state where the magnetic energy is greater than the kinetic energy. The way this can occur is through a spatial
displacement of the magnetic and velocity field structures that contain the bulk of the energy of the system, as seen in
Figure \ref{fig:overlap}. Such features can survive because of the general rotational collimation of the flow. Namely,
they avoid being disrupted by less rotationally-aligned flows, which are decreasingly prevalent in more rapidly rotating
systems. This latter process is one reason why such super-equipartition states do not appear to occur at lower rates of
rotation, for the magnetostrophic state is not achieved at those slow rates (see Figure \ref{fig:scaling}). Moreover, in
regions where these structures of fast flows and strong magnetic fields overlap and within the context of their
statistics (Figure 13), the flow and the magnetic field are nearly aligned and the Lorentz force vector is effectively
orthogonal to the velocity field. This has two effects. The first effect is to minimize the Lorentz work done on the
velocity field and the second effect is to reduce the local induction. Furthermore, the timescales for diffusion in
these simulations are sufficiently long that these structures do not dissipate quickly, and so the distributed dynamo
action achieved at intermediate scales has time to generate magnetic field that can then be advected into the
longer-lived magnetic structures. Indeed, the bulk of the inductive production of magnetic energy takes place in the
regions where the flow and magnetic field have magnitudes that are near or below their median values and at
intermediate scales. Thus the magnetic structures survive by absorbing this intermediate-scale field as it is advected
into them, which gives rise to an inverse cascade of magnetic energy.

\section{Reflections} \label{sec:conclude} 

The interaction of rotation, convection, and magnetism within the convective core of a $\Msun{10}$ B-type star has been
examined in a suite of 3D MHD simulations using the ASH code.  Those basic physical elements excite vigorous dynamo
action within the convective core captured within the simulations, where very strong megagauss magnetic fields are
established.  When magnetic fields are present, very little differential rotation is present and most of the
magnetic energy resides in the fluctuating magnetic field. The mean-field component of the magnetic field is
comparatively small, with its energy comprising at most $6\%$ of the total magnetic energy.  So, while one may be able
to characterize the evolution of the mean magnetic field in terms of an effective field theory, it is not dynamically
very relevant. However, it appears that the evolution of the mean fluctuating magnetic energy can be characterized in a
nearly equivalent way, with the primary effect being an $\alpha$-like effect.  In particular, there is a coupling
between the kinetic and current helicities that can largely account for the generation of fluctuating magnetic energy.

Given the intricate dynamics that yield angular momentum transport, magnetic induction, and Lorentz force feedback
realized in these simulations of turbulent convective core dynamos, it is striking that the clear pattern and scaling of
the ratio of the magnetic and kinetic energies seen in \S\ref{sec:scaling} emerges from them. Indeed, the
super-equipartition magnetic states are achieved once the rate of rotation is sufficiently large to induce
magnetostrophy. For these simulations, this transition occurs at about $1\%$ of breakup velocity. These dynamos can
sustain such states without catastrophic Lorentz force quenching through a spatial displacement of the magnetic and
velocity field structures. In particular, the flow and magnetic field structures formed in these simulations tend to
minimize the local Lorentz work, for the Lorentz force vector is orthogonal to the flow direction.  The local cross
helicity is maximized through the alignment of the flow and the magnetic field, which implies that the local magnetic
induction is reduced.

A tantalizing piece of observational evidence for strong dynamo states in formerly convective cores, albeit for somewhat
less massive stars, has recently been found in red giant stars. In \citet{stello16}, it is shown that a significant
fraction of moderate-mass stars that likely possessed a convective core have suppressed dipolar acoustic modes.  Those
modes can be internally reflected and trapped deep within the star and thus are observed with a greatly reduced
amplitude \citep{fuller15}.  Hence, these numerical simulations of the cores of B-type stars seem to provide additional
support for the existence of strong core dynamo-generated magnetic fields.  Yet to make greater contact with
observations of magnetic fields at the surface of massive stars, it would be useful to examine the coupling of the core
dynamo to a fossil field existing in the radiative zone within the context of the B-type stars.  However, given the
results of \citet{featherstone09} for A-type stars, one might expect super-equipartition states with even greater ratios
of magnetic to kinetic energy than seen in the simulations presented above.  However, the question about the impact of
the core dynamo on the configuration of the fossil field itself remains unanswered.

These simulations have revealed dynamo processes within the cores of massive stars that may have implications for later
stages of stellar evolution. Indeed, the powerful magnetic fields and super-equipartition states attained in these
simulations of main-sequence convective core dynamos could provide a strong-field initial condition for those more
evolved states, perhaps even providing some magnetic conditions for the pre-supernova formation of the iron core.  Since
the flow velocity within the convective core decreases and its rate of rotation increases during those later
evolutionary stages, the Rossby number is substantially lower than those realized here.  Therefore, the processes seen
to be at work in these simulations may play a role during the phases leading to the even stronger magnetic fields
established in neutron stars or magnetars during their formation.  Finally, there may be observational signatures of
these later-stage core dynamos in massive stars, given that the constraints for mode trapping require magnetic fields of
around $10^6$~G \citep{fuller15} and considering that these simulations indicate that strong fields may be achieved in
their core even during the main sequence.

\section*{Acknowledgments}

The authors thank an anonymous referee, Benjamin Brown, and Nicholas Featherstone for helpful comments. This research
and K.~C. Augustson were primarily supported by the NCAR Advanced Study Program and partly by NASA through support to
him and to J. Toomre through the Theoretical and Computational Astrophysics Network (TCAN) grant NNX14AB56G and the
Heliophysics grants NNX11AJ36 and NNX13AG18G. A.~S. Brun acknowledges financial support by the ERC through grant 207430
STARS2, and by CNRS/INSU via Programmes Nationaux Soleil-Terre and Physique Stellaire. A.~S. Brun is also grateful to
the University of Colorado and JILA for their hospitality.  The computations were primarily carried out on Pleiades at
NASA Ames with SMD grants g26133 and s0943. This work also utilized the Janus supercomputer, which is supported by the
NSF award CNS-0821794 and the University of Colorado Boulder.

\bibliography{apj-jour,bstar}

\appendix{
The strikingly different angular velocity profiles established in the convective cores of the hydrodynamic and magnetic
cases can be interpreted in terms of differing angular momentum flux balances. This is particularly true when they are
suitably averaged in time and longitude. First consider the equation for the evolution of the angular momentum density,
which can be obtained from an appropriate combination of the continuity and momentum equations of compressive MHD as

\begin{align}
  \frac{\partial \rho L}{\partial t} = -\dvg{\left[\rho\vv L + \lambda\mathbf{D} - \frac{\lambda}{4\pi}\mathbf{M}\right]} -\sddp{P_{tot}}.
\end{align}

\noindent where $L=\lambda(\vcp+\lambda\Omega_0)$, $\mathcal{M}$ is the Maxwell stress tensor, and
$P_{tot} = P_{gas} + B^2/8\pi$ is the total pressure. If a temporal and azimuthal average of this equation is taken,
then

\begin{align}
  \bigg\langle\frac{\partial \rho L}{\partial t}\bigg\rangle_{t,\phi} = -\bigg\langle\dvg{\left[\rho\vv L+\lambda\mathbf{D}
      - \frac{\lambda}{4\pi}\mathbf{M}\right]}\bigg\rangle_{t,\phi}.
\end{align}

\noindent Now consider mean fields denoted by an overbar that depend only on $r$ and $\theta$ and time-dependent
fluctuating fields with zero mean such that

\begin{align}
  \vB &= \mathbf{\overline{B}}+\mathbf{B}', \quad \vv = \mathbf{\overline{v}}+\mathbf{v}',
         \quad L = \overline{L}+L' = \left(\lambda^2\Omega_0 + \lambda\overline{v}_{\phi}\right) + \lambda v'_{\phi}, \nonumber \\
   P &= \overline{P}+P', \quad \rho = \orho+\rho'. \nonumber
\end{align}

\noindent Since $\mathcal{D}$ is a linear operator on $\vv$, and after applying the average to the expansion of these
fields, it is found that

\begin{align}
  &\boldsymbol{\nabla\cdot}\Big[\orho\mathbf{\overline{v}} \overline{L}
    + \widetilde{\rho'\mathbf{v}'}\overline{L} + \mathbf{\overline{v}}\widetilde{\rho'L'}
    + \orho\widetilde{\mathbf{v}'L'} + \widetilde{\rho'\mathbf{u}'L'} \nonumber \\
  & + \lambda\mathbf{\overline{D}} + \lambda\mathbf{D}' -
    \frac{\lambda}{4\pi}\left(\mathbf{\overline{B}_P}\overline{B}_{\phi} +
    \widetilde{\mathbf{B'_P}B'_{\phi}} \right) \Big] = 0,
\end{align}

\noindent where a tilde indicates a time and longitudinal average, the mean diffusive flux is
$\mathbf{\overline{D}}= 2\orho\nu\mathcal{D}\left[\mathbf{\overline{v}}\right]\pht$, the fluctuating
mean is $\mathbf{D}'= 2\nu\widetilde{\rho'\mathcal{D}\left[\mathbf{v}'\right]}\pht$, and
$\mathbf{B}_P$ is the poloidal field.  The anelastic approximation usually implies that
$\rho'/\orho\approx\epsilon\ll1$, where $\epsilon$ is an anelastic expansion parameter. Therefore,
the averaged compressible equation above is reduced to

\begin{align}
  \dvg{\left[\orho\mathbf{\overline{v}}\overline{L} +
      \orho\widetilde{\mathbf{v}'L'}+\lambda\mathbf{\overline{D}} -
      \frac{\lambda}{4\pi}\left(\mathbf{\overline{B}_P}\overline{B}_{\phi} +
      \widetilde{\mathbf{B'_P}B'_{\phi}} \right)\right]} \approx
  \mathcal{O}\left(\epsilon\right). \label{eqn:angmom}
\end{align}

Since the general symmetry of the flow and magnetic field in the core is cylindrical, it is useful to focus on the
cylindrical components of the angular momentum flux. In particular, given Equation (\ref{eqn:angmom}), it is now easy to
identify each of the vertically averaged angular momentum fluxes in the cylindrically radial direction as

\begin{align}
  \langle F_\lambda^{MC}\rangle_z &= \frac{1}{2\lambda}\int_{-\lambda}^{\lambda} dz \orho\overline{v}_\lambda\overline{L},
  \quad \langle F_\lambda^{RS}\rangle_z = \frac{1}{2\lambda} \int_{-\lambda}^{\lambda} dz \orho\widetilde{v'_\lambda L'}, \nonumber \\
  \langle F_\lambda^{MM}\rangle_z &= \frac{1}{8\pi} \int_{-\lambda}^{\lambda} dz \overline{B}_\lambda\overline{B}_{\phi},
  \quad \langle F_\lambda^{FM}\rangle_z = \frac{1}{8\pi} \int_{-\lambda}^{\lambda} dz \widetilde{B'_\lambda B'_{\phi}}, \nonumber \\
  \langle F_\lambda^{VS}\rangle_z &= \frac{1}{2} \int_{-\lambda}^{\lambda} dz \mathbf{\overline{D}}\cdot\hat{\lambda}, \label{eqn:angmomflux}
\end{align}

\noindent where
$\langle F_\lambda^{tot}\rangle_z = \langle F_\lambda^{MC}\rangle_z + \langle F_\lambda^{RS}\rangle_z + \langle
F_\lambda^{MM}\rangle_z + \langle F_\lambda^{FM}\rangle_z + \langle F_\lambda^{VS}\rangle_z$.
These fluxes are shown for cases H4 and M4 in Figures \ref{fig:hdvmhd}(c) and (f).}

\end{document}